\documentclass[useAMS,usenatbib]{mn2e}

\usepackage{epsfig,color,pifont,array}
\usepackage{wasysym}

\DeclareMathAlphabet{\mathpzc}{OT1}{pzc}{m}{it}

\usepackage{natbib}
\bibpunct{(}{)}{;}{a}{}{,}
\def\d{{\rm{d}}}
\def\ts{{\thinspace}}
\def\lb{{\langle}}
\def\rb{{\rangle}}
\def\llb{\left\langle}
\def\rrb{\right\rangle}

\def\spose#1{\hbox to 0pt{#1\hss}}
\def\ltsimm{\mathrel{\spose{\lower 3pt\hbox{$\sim$}}
        \raise 2.0pt\hbox{$<$}}}
\def\gtsimm{\mathrel{\spose{\lower 3pt\hbox{$\sim$}}
        \raise 2.0pt\hbox{$>$}}}

\def\cm{{\rm\thinspace cm}}

\def\s{{\rm\thinspace s}}

\def\g{{\rm\thinspace g}}

\def\erg{{\rm\thinspace erg}}
\def\Hz{{\rm\thinspace Hz}}

\def\ster{{\rm\thinspace ster}}
\def\ergps{\hbox{${\rm\erg\s^{-1}\,}$}}

\def\pcm{\hbox{${\rm\cm^{-1}\,}$}}
\def\pcm2{\hbox{${\rm\cm^{-2}\,}$}}
\def\pcm3{\hbox{${\rm\cm^{-3}\,}$}}
\def\ergpscm3Hz{\hbox{${\rm\ergps\cm^{-3}\Hz^{-1}\,}$}}
\def\ergpscm3Hzster{\hbox{${\rm\ergps\cm^{-3}\Hz^{-1}\ster^{-1}\,}$}}
\def\gpcm3{\hbox{${\rm\g\cm^{-3}\,}$}}
\def\ergpcm2{\hbox{${\rm\erg\cm^{-2}\,}$}}
\def\ergpcm3{\hbox{${\rm\erg\cm^{-3}\,}$}}
\def\phpscm2{\hbox{${\rm photons\s^{-1}\cm^{-2}\,}$}}

\def\aap{{\rm A\&A}}
\def\apj{{\rm ApJ}}
\def\apjl{{\rm ApJL}}
\def\apjs{{\rm ApJS}}

\def\mnras{{\rm MNRAS}}
\def\nat{{\rm Nature}}
\def\apss{{\rm Ap\&SS}}
\def\araa{{\rm ARA\&A}}

\def\jcp{{\rm J.~Comput.~Phys}}

\def\nar{{\rm NewAR}}
\def\pasj{{\rm PASJ}}

\title [Global MHD disks III]{Global simulations of magnetorotational
  turbulence III: influence of field configuration and mass injection}

\author[E.~R.~Parkin]
{E.~R.~Parkin\thanks{E-mail: ross.parkin@anu.edu.au} 
  \\Research School of Astronomy and Astrophysics,
  Australian National University, Canberra, ACT 2611, Australia}

\begin{document}

\date{Accepted ... Received ...; in original form ...}

\pagerange{\pageref{firstpage}--\pageref{lastpage}} \pubyear{2013}

\maketitle

\label{firstpage}

\begin{abstract}
  {The stresses produced by magnetorotational turbulence can provide
    effective angular momentum transport in accretion disks. However,
    questions remain about the ability of simulated disks to reproduce
    observationally inferred stress-to-gas-pressure ratios. In this
    paper we present a set of high resolution global
    magnetohydrodynamic disk simulations which are initialised with
    different field configurations: purely toroidal, vertical field
    lines, and nested poloidal loops. A mass source term is included
    which allows the total disk mass to equilibrate in simulations
    with long run times, and also enables the impact of rapid mass
    injection to be explored. Notably different levels of angular
    momentum transport are observed during the early-time transient
    disk evolution. However, given sufficient time to relax, the
    different models evolve to a statistically similar quasi-steady
    state with a stress-to-gas-pressure ratio, $\lb \alpha_{\rm P} \rb
    \sim 0.032-0.036$. Such behaviour is anticipated based on
    consideration of mean magnetic field evolution subject to our
    adopted simulation boundary conditions.

    The indication from our results is that {\it steady, isolated}
    disks may be unable to maintain a large-scale magnetic field or
    produce values for the stress-to-gas-pressure ratio implied by
    some observations. Supplementary simulations exploring the
    influence of trapping magnetic field, injecting vertical
      field, and rapidly injecting additional mass into the disk show
      that large stresses can be induced by these mechanisms. In the
      first instance, a highly magnetized disk is produced with $\lb
      \alpha_{\rm P} \rb \sim 0.21$, whereas the latter cases lead to
      a transient burst of accretion with a peak $\lb \alpha_{\rm P}
      \rb \simeq 0.1-0.25$. As a whole, the simulations highlight the
      common late-time evolution and characteristics of turbulent
      disks for which the magnetic field is allowed to evolve freely
      (i.e., without constraint/replenishment). In contrast, if the
      boundaries of the disk, the rate of injection of magnetic field,
      or the rate of mass replenishment are modified to mimic
      astrophysical disks, markedly different disk evolution occurs.
  }

\end{abstract}




\section{Introduction}
\label{sec:intro}

For disk accretion to occur angular momentum must be removed from
orbiting material. It has long been thought that magnetic fields could
play more than just a passive role in this situation
\citep{Shakura:1973, Lynden-Bell:1974}. The magnetorotational
instability \citep[MRI -][]{BH91,BH92,BH98} has since emerged as a
robust mechanism for destabilising a differentially rotating disk,
placing the importance of magnetic fields on a firm
footing. Subsequent numerical work has demonstrated that the
non-linear motions induced by the MRI lead to self-sustaining
turbulence with the resulting stresses effectively transporting
angular momentum \citep[e.g.][]{Hawley:1995, Brandenburg:1995,
  Stone:1996}.

A large body of the numerical work on magnetorotational turbulence
to-date has focused on a local patch of disk with the vertical
component of gravity neglected - the unstratified
shearing-box. Studies examining various magnetic field configurations
have produced a wide range of values for the stress-to-gas-pressure
ratio, $\lb \alpha_{\rm P} \rb \simeq 0.001- 0.3$ \citep[e.g.][to
mention but a few]{Hawley:1995, Sano:2004, Fromang:2007a, Simon:2009,
  Guan:2009}. Values of $\lb \alpha_{\rm P} \rb \gtsimm 0.01$ are,
however, only achieved by models initialised with a net vertical
magnetic field. By construction, the unstratified shearing-box
preserves an initial net vertical field. Hence, in these cases power
can be effectively injected by vertical MRI modes on large spatial
scales, feeding the turbulent cascade from the top down.

Astrophysical accretion disks are, however, intrinsically global
objects (by which we mean they are not periodic in the radial and/or
vertical directions). Recent studies by \cite{Sorathia:2010,
  Sorathia:2012} and \cite{Beckwith:2011} have demonstrated that
localised patches of global disks do not preserve their magnetic field
configuration in the same manner as an unstratified
shearing-box. Moreover, an initial vertical field is free to evolve in
global models, and it is not clear whether global disks will achieve
the large stresses implied by net-vertical flux unstratified
shearing-box models. What dependence of the turbulent stresses on
initial field configuration should one expect for a stratified global
disk?  Newtonian global disk models exhibit $\lb \alpha_{\rm P} \rb
\sim 0.001-0.15$ \citep[e.g.][]{Hawley:2000, Fromang:2006, Flock:2011,
  Beckwith:2011, Hawley:2011, Hawley:2013, Parkin:2013, Parkin:2013b,
 Gressel:2013, Parkin:2014, Suzuki:2014}. Larger values of $\lb
\alpha_{\rm P} \rb$ are typically associated with the saturation
stress level during the initial transient phase of the simulation,
with some evidence for stronger stresses when the disk initially
possesses a poloidal magnetic field. There are mixed results from
general relativistic global disk simulations. For instance,
\cite{Beckwith:2008} find markedly similar disk properties
irrespective of the initial field configuration. In contrast,
\cite{McKinney:2012} \citep[see also -][]{Tchekhovskoy:2011} quote
$\lb \alpha_{\rm P} \rb \sim 0.01-0.69$ with the lowest values for an
initially toroidal field, whereas the largest values are for models
where the disk initially contains a large quantity of poloidal
magnetic flux.

In the present study we investigate the dependence of
magnetorotational turbulence in a global disk on the initial field
configuration. The simulations include a mass source term which
facilitates simulations with long run times. Thus, the late-time
quasi-steady disk evolution can be studied without concerns about
contamination from mass exhaustion. This differs from previous global
disk models, with the exception of \cite{Flock:2011, Flock:2012,
  Flock:2012b} and \cite{Parkin:2014}. A key result is the ignorance
of the quasi-steady state turbulent stress, and thus efficiency of
angular momentum transport, to the initial field configuration; given
sufficient time the disk expels and/or digests the initial field,
evolving to a statistically almost identical state. The simulations
produce $\lb \alpha_{\rm P} \rb \sim 0.032-0.036$ during the
quasi-steady state, which falls short of observationally inferred
values of $\sim 0.1-0.4$ \citep[see][and references
therein]{King:2007}. We perform supplementary simulations to address
this quandary, finding that $\lb \alpha_{\rm P} \rb \sim 0.2$ can be
achieved by trapping field, and $\lb \alpha_{\rm P} \rb \sim 0.1-0.25$
results when mass or vertical magnetic field are injected. The former
case is relevant for magnetically arrested \citep{Narayan:2003} and
magnetically levitating \citep{Johansen:2008, Gaburov:2012} disk
models.

The remainder of this paper is organised as follows: the simulation
setup and initial conditions are described in \S~\ref{sec:model}. A
preparatory discussion of the evolution of mean magnetic fields
subject to boundary conditions is given in \S~\ref{sec:mean_field},
followed by the results of the simulations in \S~\ref{sec:results}. A
discussion of the implications of the results for astrophysical disks,
and some potentially interesting avenues for future studies, are
discussed in \S~\ref{sec:discussion}. Finally, we close with
conclusions in \S~\ref{sec:conclusions}.

\section{The model}
\label{sec:model}

\subsection{Simulation code}
\label{subsec:hydromodel}

The global disk simulations are performed by solving the
time-dependent equations of ideal magnetohydrodynamics (MHD) using the
{\sevensize PLUTO} code \citep{Mignone:2007} in a 3D spherical
$(r,\theta,\phi)$ coordinate system. The grid used for all simulations
in this work has ($n_{r},n_{\theta},n_{\phi}$) = (512,\ts 256,\ts 256)
uniformly spaced cells, covering the spatial extent $8<r<34$, $\pi/2 -
\theta_{0} <\theta< \pi/2 + \theta_{0}$ (where $\theta_{0}=\tan^{-1}(3
H/R )$), and $0< \phi < \pi /2$. (The $\theta$-extent of the grid
equates to $|z|\pm 3H$ for a constant aspect ratio disk.) The
simulated disks all have $H/R=0.1$, where $H$ is the disk scale
height. In terms of cells per scale height, the adopted grid has
($n_{r}/H,n_{\theta}/H,n_{\phi}/H) \simeq (16-67, 44, 16)$ which is
comfortably within the regime of convergence with resolution detailed
by \cite{Parkin:2013b}. The simulations are performed in scaled units:
$\rho_{\rm scale}=1.67\times10^{-7}~{\rm gm~s^{-1}}$, $v_{\rm
  scale}=c$ (where $c$ is the speed of light), $T_{\rm scale}=\mu m
c^{2}/k_{\rm B} = 6.5\times10^{12}~{\rm K}$, $l_{\rm
  scale}=1.48\times10^{13}~{\rm cm}$, and the value of $l_{\rm scale}$
corresponds to the gravitational radius of a $10^{8}~{\rm M_{\odot}}$
black hole.

The adopted numerical setup closely follows that of
\cite{Parkin:2014} - a detailed description can be found there-in. In
brief, {\sevensize PLUTO} code was configured to use the five-wave
HLLD Riemann solver of \cite{Miyoshi:2005}, piece-wise parabolic
reconstruction \citep[PPM -][]{Colella:1984}, limiting during
reconstruction on characteristic variables \citep[e.g.][]{Rider:2007},
second-order Runge-Kutta time-stepping, the upwind Constrained
Transport scheme \citep{Gardiner:2005}, and the FARGO-MHD module
\citep[][]{Mignone:2012}. A mass source term is included which relaxes
the gas density in the region $31 \leq r \leq 34$, $|z|<2H$ towards
the initial density distribution over a timescale of an orbital
period, where $H$ is the thermal disk scale-height. This latter
addition allows the total mass and energies in the disk to stabilise,
permitting long run-time global disk simulations which avoid the
restriction of exhausting the mass supply \citep[][]{Flock:2011,
  Parkin:2014}.

The adopted boundary conditions are identical to those used in
\cite{Parkin:2014}, with the exception that for model VERT-TRAP the
radial velocity at the inner radial boundary is set to $v_{r} = -(3/2)
\alpha_{\rm P} c_{\rm s}^{2}/\Omega r$, where $c_{\rm s}$ is the sound
speed, $\Omega$ is the angular velocity, and $\alpha_{\rm P}$ is the
\cite{Shakura:1973} $\alpha$-parameter (see Eq~\ref{eqn:alphaP}). This
equates to fixing the radial velocity to a viscous outflow rate and
has been used previously by \cite{Fromang:2006} and \cite{Suzuki:2014}
to reduce the rate at which mass exits the grid. In the present study
this boundary condition is utilised to deliberately trap magnetic
field, by adopting $\alpha_{\rm P}= 0.001$ in the equation for $v_{r}$
(which is lower than the $\lb \alpha_{\rm P} \rb= 0.032-0.036$
produced by the turbulent disk).

\begin{table}
\begin{center}
  \caption[]{Summary of simulations. $\beta_{0}$ is the plasma-$\beta$
    of the initial magnetic field.} \label{tab:simlist}
\begin{tabular}{llll}
  \hline
  Model & Initial field & $\beta_0$ & Comment \\
  \hline
  TOR-$\beta20$ & Toroidal & 20 & Fiducial model \\
  TOR-$\beta1$ & Toroidal & 1 & Low $\beta_{0}$ \\
  VERT & Vertical & 2000 & Vertical field \\
  LOOP & Poloidal loops & 300 & Zero-net flux field \\
  VERT-TRAP & Vertical & 2000 & Restricted radial outflow \\
  VERT-B+ & Vertical & 2000 & Testing $B_{\theta}$ injection \\
  TOR-M+ & Toroidal & 20 & Testing mass injection \\
  \hline
\end{tabular}
\end{center}
\end{table}

\subsection{Outline of simulations performed}
Simulations are performed to examine the following topics (details are
summarised in Table~\ref{tab:simlist}):
\begin{itemize}
\item {\it Initial field configuration}: The primary aim of this study
  is to investigate the influence of different initial magnetic field
  topologies on the subsequent non-linear evolution of turbulence in a
  disk. To this end we examine three different initial field
  topologies: i) a purely toroidal magnetic field (models
  TOR-$\beta20$ and TOR-$\beta1$), ii) purely vertical magnetic field
  (model VERT), and, iii) concentric nested loops of poloidal magnetic
  field (model LOOP).
\item {\it Strongly magnetized disk}: Model TOR-$\beta1$ examines the
  affect on both early- and late-time disk evolution associated with a
  strong initial magnetic.
\item {\it Trapping of magnetic field}: Model VERT-TRAP is a repeat of
  model VERT with an intentionally restrictive outflow boundary
  condition used at the inner radial boundary, allowing the influence
  of field trapping on the disk evolution to be examined.
\item {\it Injection of vertical magnetic field}: Model VERT-B+
    explores the disk evolution when a vertical magnetic field is
    steadily added to the simulation domain.
\item {\it Transient accretion induced by rapid mass injection}: A
  supplementary simulation (model TOR-M+) has been performed for
  which the mass source term (described in \S~\ref{subsec:hydromodel})
  is modified to inject mass rapidly over a short duration of time,
  resulting in a transient burst in accretion activity in the
  disk. Further details of model TOR-M+ are given in
  \S~\ref{subsec:tor-inj}.
\end{itemize}

\subsection{Initial conditions}
\label{subsec:initialconditions}

For models TOR-$\beta20$ and TOR-$\beta1$ we use the exact equilibrium
torus of \cite{Parkin:2013} with the ratio of gas-to-magnetic
pressure, $\beta = 2p/|B|^2 \equiv 2p/B_{\phi}^2$ initially set to 20
and 1, respectively. The magnetic field is net-flux and purely
toroidal, with a constant $\beta$ throughout the disk. The density
distribution of this disk in cylindrical ($R,z$) coordinates is given
by,
\begin{equation}
  \rho(R,z) = \rho(R,0) \exp\left(\frac{-(\Phi(R,z) - \Phi(R,0))}{T(R)} \frac{\beta}{1 + \beta} \right), \label{eqn:rho}
\end{equation}
\noindent where $p = \rho T$, and $T=T(R)$ is the (isothermal in
height) temperature. For the radial profiles $\rho(R,0)$ and $T(R)$ in
Eq~(\ref{eqn:rho}) we use simple functions inspired by the
\cite{Shakura:1973} disk model:
\begin{eqnarray}
  \rho(R,0) & = & \rho_{0} f(R,R_{0},R_{\rm out})
  \left(\frac{R}{R_{0}}\right)^{\epsilon}, \\
  T (R) & = & T_{\rm 0} \left(\frac{R}{R_{0}}\right)^{\eta}, \label{eqn:temp}
\end{eqnarray}
\noindent where $\rho_{0}$ sets the density scale, $R_{0}$ and $R_{\rm
  out}$ are the radius of the inner and outer disk edge, respectively,
$f(R,R_{0},R_{\rm out})$ is a tapering function that truncates the
density profile at a specified inner and outer radius \citep{Parkin:2013},
and $\epsilon$ and $\eta$ set the slope of the density and temperature
profiles, respectively. Values are set to $R_{0}=7$, $R_{\rm out}=50$,
$\rho_{0}=10$, $\epsilon=-33/20$, $\eta=-9/10$, and
$T_{0}=1.5\times10^{-3}$, producing a disk with $H/R=0.1$. The
rotational velocity of the disk is,
\begin{eqnarray}
  v_{\phi}^2(R,z) = v_{\phi}^2(R,0) + (\Phi(R,z) -
  \Phi(R,0)) \frac{R}{T}\frac{d T}{d R}, \label{eqn:vphi}
\end{eqnarray}
\noindent where,
\begin{eqnarray}
v_{\phi}^2(R,0) = R \frac{\partial
  \Phi (R,0)}{\partial R} + \frac{2 T}{\beta} + \nonumber \\
 \left(\frac{1 + \beta}{\beta} \right)\left(\frac{R
    T}{\rho(R,0)}\frac{\partial \rho(R,0)}{\partial R}
 + R \frac{d T}{d R}\right).   \label{eqn:vphi0}
\end{eqnarray}
The (purely toroidal) magnetic field is initialised using the
$\theta$-component of the vector potential,
\begin{equation} 
  A_{\theta} = \frac{1}{r} \int_{r_{0}}^{r} r B_{\phi} \d r, 
\end{equation}
where $r_{0} = 7$ and $A_{\theta}(r \leq r_{0})=0$. The reader is
referred to \cite{Parkin:2013} and \cite{Parkin:2014} for further
details of the initial conditions for model TOR-$\beta20$.

For the poloidal field models (VERT, LOOP, VERT-TRAP, and VERT-B+) the
initial conditions consist of an isothermal-in-height disk ($T=T(R)$)
which is in hydrodynamic equilibrium. The equations for the density
and rotational velocity in this case are found from the $\beta
\rightarrow \infty$ limit of Eqs~(\ref{eqn:rho}) and
(\ref{eqn:vphi0}). The magnetic field is initialised in the poloidal
field models using the $\phi$-component of the vector potential. For
models VERT, VERT-TRAP, and VERT-B+ we use,
\begin{equation}
A_{\phi} \propto \frac{1}{2R} (R^2 - R_{0}^2), \label{eqn:Aphi1}
\end{equation}
with $R_{0}=7$. The value of $A_{\phi}$ in Eq~(\ref{eqn:Aphi1}) is
scaled to give a disk averaged initial field strength with $\beta
\simeq 2000$. For model LOOP we initialise the magnetic field via:
\begin{equation}
\begin{array}{l l l}
A_{\phi}  \propto & (\rho/\rho_0) \sin (k_{\rm loop} (R-R_1)); & R_1 < R < R_2, \nonumber\\
 A_{\phi} = & 0; &  {\rm otherwise,} \label{eqn:Aphi2} 
\end{array}
\end{equation}
where $\rho_0$ is the disk density at ($R=R_{0},z=0$), $k_{\rm loop}=
\pi n_{\rm loop}/(R_2 -R_1)$, $n_{\rm loop}=10$ is the number of
loops, $R_0=7$, $R_1=10$, and $R_2=30$. This produces a zero-net
poloidal flux field consisting of ten concentric loops between
$10<R<30$. Equation~(\ref{eqn:Aphi2}) is scaled to give a
disk-averaged $\beta \simeq 300$.

In all models, we initiate the development of turbulence by adding
random poloidal velocity fluctuations of amplitude $0.01~c_{\rm s}$ to
the initial equilibrium.

\section{Mean field evolution}
\label{sec:mean_field}

Three-dimensional power spectra computed from turbulent global disk
models show that the majority of energy is contained in scales on the
order of $\sim 2 H$ \citep{Parkin:2013b, Parkin:2014}. In such a case
volume-averaged mean fields are a proxy for large-scale fields, which
follows from the integral transform relation between length scales and
wavenumber space (i.e. Parseval's theorem). Moreover, the injection of
energy into the turbulent cascade on large scales requires large-scale
field components, so that tracking the evolution of mean magnetic
fields provides information about the ability of the field
configuration to inject energy at the upper end of the cascade. The
simulation boundary conditions influence the evolution of mean
magnetic fields, and in the following we examine the consequences of
our adopted setup. (For further discussion of these points, and the
associated impact on the convergence properties of different numerical
setups, see \citeauthor{Parkin:2013b}~\citeyear{Parkin:2013b} and
references therein.)

Let us consider the volume averaged induction equation for the radial,
vertical ($\theta$-direction), and azimuthal magnetic fields,
respectively,
\begin{eqnarray}
\frac{\partial}{\partial t} \lb B_{r} \rb = 
 -\frac{1}{V}\int_{\theta_2} \mathcal{E}_{\phi} \ts \d s_{\theta} 
+ \frac{1}{V} \int_{\theta_1} \mathcal{E}_{\phi} \ts \d s_{\theta},   \label{eqn:br_surf}
\end{eqnarray}
\begin{eqnarray}
\frac{\partial}{\partial t} \llb \frac{B_{\theta}}{r } \rrb  = 
 \frac{1}{V}\int_{r_2}  \frac{\mathcal{E}_{\phi}}{r} \d s_{r} 
- \frac{1}{V} \int_{r_1}  \frac{\mathcal{E}_{\phi}}{r} \d s_{r}  \label{eqn:bth_surf}
\end{eqnarray}
and,
\begin{eqnarray}
\frac{\partial}{\partial t} \llb \frac{B_{\phi}}{r \sin \theta} \rrb  = 
 -\frac{1}{V}\int_{r_2} \frac{\mathcal{E}_{\theta}}{r \sin \theta} \d s_{r} 
+ \frac{1}{V} \int_{r_1} \frac{\mathcal{E}_{\theta}}{r \sin \theta} \d s_{r}  \nonumber \\
-\frac{1}{V}\int_{\theta_2} \frac{\mathcal{E}_{r}}{r \sin \theta} \d s_{\theta} 
+ \frac{1}{V} \int_{\theta_1} \frac{\mathcal{E}_{r}}{r \sin \theta}\d s_{\theta} \label{eqn:bphi_surf}
\end{eqnarray}
where $V = \iiint dV$ is the volume bound by the surfaces, $\d s_{r} =
r^2 \sin \theta \ts \d \theta \ts \d \phi$, $\d s_{\theta} = r \sin
\theta \ts \d r \ts \d \phi$. The $\theta$- and $\phi$-components of
the electric field are given by $\mathcal{E}_{\theta}=v_{r} B_{\phi} -
v_{\phi} B_{r}$ and $\mathcal{E}_{\phi} = v_{\theta} B_r - v_r
B_{\theta}$, respectively. Angle-brackets denote a volume-average,
e.g.,
\begin{equation}
  \langle q\rangle = \frac{\iiint q \ts r^2 \sin \theta \d r \ts  \d
    \theta \ts \d\phi}{\iiint r^2 \sin
    \theta \ts \d r \ts \d\theta \ts d\phi}.
\end{equation}

From Eq~(\ref{eqn:bth_surf}) one sees that the vertical field can
evolve when the radial boundaries are open, which is the case for the
global disk simulations in this work. In contrast, the initial mean
vertical magnetic field is preserved in shearing-box studies employing
a radially periodic boundary condition - see \cite{Hawley:1995} for
further details. Shearing-box models do not, therefore, allow one to
investigate questions of if/how a vertical magnetic field is preserved
because the vertical field evolution is constrained by the simulation
setup. Vertical field can naturally disperse through radial boundaries
in the global models considered in this work, and the tendency in the
simulations is for the initial field configuration to advect off the
grid in the accretion flow.

Considering Eqs~(\ref{eqn:bth_surf}) and (\ref{eqn:bphi_surf}) in more
detail, the rate of dispersal of the initial field configuration can
be limited by, for example, restricting the radial outflow velocity at
the boundaries because $v_{r}$ features in both $\mathcal{E}_{\theta}$
and $\mathcal{E}_{\phi}$. This effect is explored with model VERT-TRAP
in \S~\ref{subsec:vert-vbc}, where it is found that trapping vertical
and azimuthal magnetic field leads to drastically different disk
evolution. Alternatively, the mean vertical field strength may be
altered by adding magnetic field to the domain, which is explored
using model VERT-B+ in \S~\ref{subsec:vert-Bthadd}.

\section{Results}
\label{sec:results}

\begin{figure}
  \begin{center}
    \begin{tabular}{c}
\resizebox{80mm}{!}{\includegraphics{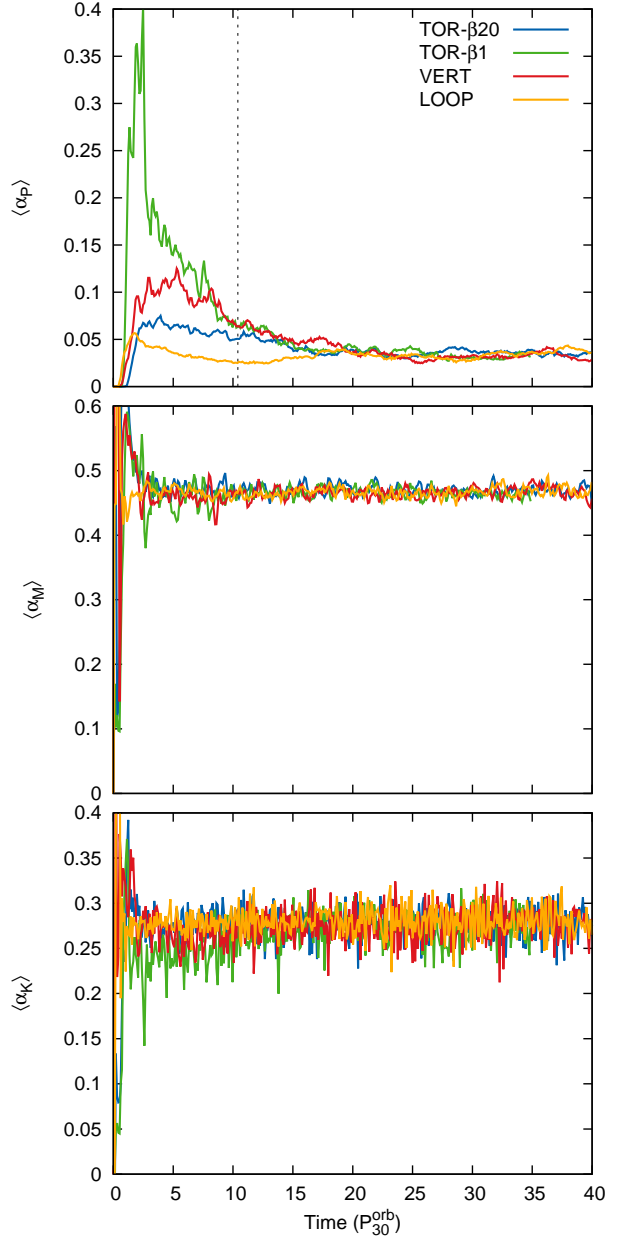}} \\
   \end{tabular}
   \caption{The time evolution of $\langle\alpha_{\rm P}\rangle$
     (upper), $\langle\alpha_{\rm M}\rangle$ (middle), and
     $\langle\alpha_{\rm K}\rangle$ (lower) computed for the disk body
     ($15<r<25$, $|z|<2H$). Time is in units of the orbital period at
     a radius of $r=30$, $P^{\rm orb}_{30}$. (For comparison, $P^{\rm
       orb}_{30}= 9~P^{\rm orb}_{8}$, therefore roughly 360 inner disk
     orbits are covered.) Corresponding time averaged results are
     given in Table~\ref{tab:models}. The vertical dashed line in the
     upper panel indicates the simulation run time for the models
     presented by \citeauthor{Hawley:2013}~(\citeyear{Hawley:2013}).}
    \label{fig:alpha}
  \end{center}
\end{figure}

\begin{table*}
\begin{center}
  \caption[]{List of time averaged quantities computed for the
    simulations. For all models values are spatially averaged within
    the disk body ($15<r<25$, $|z|<2H$) and time averaged over the
    interval $20 \leq t \leq 40\;P_{30}^{\rm orb}$, except for model
    TOR-$\beta1$ where the time average is between $20 \leq t \leq
    35\;P_{30}^{\rm orb}$.} \label{tab:models}
\begin{tabular}{llllll}
  \hline
  Parameter & TOR-$\beta20$ & TOR-$\beta1$ & VERT & LOOP & VERT-TRAP \\
  \hline
  $M_{0}$ & 5477 & 6759 & 5376 & 5376 & 5376 \\
  $M_{\rm QSS}$ & 1814 & 1779 & 1870 & 2118 & 424 \\
  $\langle\alpha_{\rm P}\rangle$ & 0.036 &  0.034 & 0.032 & 0.034 & 0.21 \\
  $\langle\alpha_{\rm M}\rangle$ & 0.469 &  0.466 & 0.467 & 0.467 & 0.466 \\
  $\langle\alpha_{\rm K}\rangle$ & 0.279 &  0.273 & 0.278 & 0.280 & 0.176 \\
  $-\langle B'_{r}B'_{\phi}\rangle / \lb \rho v'_{r}v'_{\phi} \rb$ &
  2.69 &  2.55 & 2.55 & 2.63 & 5.71 \\
  $\dot{M}_{\rm rad}$ ($P_{30}^{\rm orb-1}$) & 160 & 156 & 159 & 154 &
  260 \\
  $\dot{M}_{\rm vert}$  ($P_{30}^{\rm orb-1}$) & 23 & 29 & 18 & 22 & 154
  \\
  $N_{\rm r}$ &  0.76 & 0.76 & 0.74  & 0.75 & 0.95 \\
  $N_{\theta}$ & 0.65  &  0.64 & 0.62  & 0.64 & 0.93 \\
  $N_{\phi}$ & 0.81 & 0.80 & 0.79 & 0.80 & 0.96 \\
  $\langle \beta_{\rm r}\rangle$ & 118 & 129 & 139 & 131 & 10 \\
  $\langle\beta_{\theta}\rangle$ & 316 & 340 & 369 & 347 &  18 \\
  $\langle\beta_{\phi}\rangle$ & 20 & 21 & 22 & 21 & 1.8 \\
  $\langle\beta_{\rm tot}\rangle$ & 16 & 17 & 18 & 17 & 1.4 \\
  $\langle B_{r}\rangle$ ($\times 10 ^{-3}$) & 0.002 $\pm$ 0.004 & -0.001$\pm$0.005 & 0.001
  $\pm$ 0.003      & 0.003 $\pm$ 0.004 & -0.098 $\pm$ 0.015 \\
  $\langle B_{\theta}\rangle$ ($\times 10 ^{-3}$) & 0.033 $\pm$ 0.001
  & 0.0002 $\pm$ 0.0019 &
  -0.012 $\pm$ 0.001 & 0.006 $\pm$ 0.002 & -0.13 $\pm$ 0.01 \\
  $\langle B_{\phi}\rangle$ ($\times 10 ^{-3}$) & 0.090 $\pm$ 0.071 &
  0.14$\pm$0.09 & 0.036 $\pm$ 0.063 & -0.25 $\pm$ 0.06 & 2.86 $\pm$ 0.52 \\
  \hline
\end{tabular}
\end{center}
\end{table*}

\subsection{Influence of initial field topology in isolated disks}
\label{subsec:field_topology}

In this section we begin our investigation by considering models with
different initial field configurations and their respective evolution
when allowed to evolve freely. In this regard models TOR-$\beta1$,
TOR-$\beta20$, VERT, and LOOP represent isolated accretion disks with
no external influence and a limited source of magnetic field. In
\S~\ref{subsec:vert-vbc}, \ref{subsec:vert-Bthadd}, and
\ref{subsec:tor-inj} we consider models with modified boundary
conditions, magnetic field injection, and mass injection. These
additional models are geared towards exploring possible external
influences on the evolution of the disk turbulence.

\subsubsection{Turbulent stresses}

The simulations performed in this work aim to explore the differences
in the ensuing turbulent state arising from different initial field
topology. The models TOR-$\beta20$, TOR-$\beta1$, VERT, and LOOP start
from an almost identical disk (in terms of density, temperature, and
rotation velocity distributions) but differ in the initial strength
and topology of the magnetic field\footnote{Model TOR-$\beta20$ was
  previously presented by \cite{Parkin:2014} in the context of
  turbulent energetics and further detailed analysis of this
  simulation may be found therein.}. As the simulations evolve, the
perturbations introduced into the initial conditions excite the
MRI. Because of the different field topologies and field strengths in
the models the respective initial MRI growth rates
differ. Consequently, the resultant stresses arising during the
initial transient disk evolution are not the same between the
simulations. An indication of the differences in the turbulent
stresses during the initial transient phase can be gained from the
\cite{Shakura:1973} $\alpha$-parameter, which we define as the
$r-\phi$ component of the combined Reynolds and Maxwell stress
normalised by the gas pressure,
\begin{equation}
  \langle \alpha_{P}\rangle = \frac{\langle
    \rho v'_{r}v'_{\phi} - B'_{r}B'_{\phi}\rangle}{\langle p\rangle}, \label{eqn:alphaP}
\end{equation}
where primes denote a fluctuating (turbulent) deviation from an
azimuthal mean, such that,
\begin{equation}
\begin{array}{ c c c}
  v_{\rm i} = v'_{i} + \bar{v}_{i}, &{\rm where},
  &\bar{v}_{i} = [\rho v_{i}]/[\rho],
\end{array}
\end{equation}
and,
\begin{equation}
\begin{array}{ c c c}
  B_{i} = B'_{\rm i} + \bar{B}_{i}, &{\rm where},
  &\bar{B}_{i} = [B_{i}],
\end{array}
\end{equation}
where over-bars indicate a mean and square brackets indicate an
azimuthal average,
\begin{equation}
[q] = \frac{\int q r \sin\theta d\phi}{ \int r \sin \theta d\phi}.
\end{equation}
Note that we use density-weighted averages to compute the mean
velocities, consistent with the analytical approach of
\cite{Kuncic:2004} - see also \cite{Parkin:2014}. The volume averages
in Eq~(\ref{eqn:alphaP}), and in all other instances in this work, are
taken over the region $15<r<25$, $|z|<2H$, and the entire azimuthal
extent of the domain, $0<\phi<\pi/2$.

Fig.~\ref{fig:alpha} shows that over the first $10 \ts P_{30}^{\rm
  orb}$ of the simulations (where $P^{\rm orb}_{30}$ corresponds to
the orbital period at a radius, $r=30$) model TOR-$\beta1$ has the
highest $\lb \alpha_{\rm P} \rb$, followed by VERT, TOR-$\beta20$, and
then LOOP. The prominent spike in $\lb \alpha_{\rm P} \rb$ for model
TOR-$\beta1$ is associated with the rapid development of large-scale,
non-axisymmetric MRI modes, powered by the strong toroidal field. In
contrast, the relatively large transient stresses for model VERT
result from vertical MRI mode growth, with an additional contribution
from parasitic instabilities feeding on channel flows
\citep{Goodman:1994}.

The dependence of the turbulent stress on initial field topology was
recently examined by \cite{Hawley:2013} using high resolution global
disk simulations. Given the similarity in their goals to the present
study, a comparison is warranted. However, we note that the
simulations presented by \cite{Hawley:2013} were only run until a time
of $t\simeq 10.4 \ts P^{\rm orb}_{30}$ (indicated by the vertical
dashed line in the upper panel of Fig.~\ref{fig:alpha}) which, as
noted by the authors, was the result of the limitations of a finite
disk mass. Within this time interval - which covers the transient
evolution phase close to the beginning of the simulation - our results
agree with \cite{Hawley:2013} in that there is a clear dependence of
the peak saturation stress level on the initial field
topology. However, there are some differences in the details. For
instance, \cite{Hawley:2013} found that a higher saturation stress was
achieved by a simulation with two poloidal loops compared to one with
a toroidal field. In contrast, we find the highest saturation stress
for a toroidal field model with $\beta =1$ initially (TOR-$\beta1$),
with the lowest stress level for a model with ten concentric poloidal
loops (model LOOP).

Focusing on the latter half of the simulations in
Fig.~\ref{fig:alpha}, one sees that the turbulence evolves to be
markedly similar in the different models. Indeed, time-averaged values
are in the range $\lb \alpha_{\rm P}\rb = 0.032-0.036$ (noted in
Table~\ref{tab:models}) which is consistent with the converged
quasi-steady state found by \cite{Parkin:2013b} for a toroidal field
disk similar to TOR-$\beta20$. Importantly, this demonstrates that
given sufficient time an isolated turbulent accretion disk evolves to
a state that is identical, irrespective of the initial magnetic field
configuration. Thus, we concur with \cite{Hawley:2013} that the
turbulent stress {\it during the initial transient phase} depends on
the initial field configuration, whereas our results show that the
late-time quasi-steady stresses (i.e. at $t> 20 \ts P^{\rm orb}_{30}$)
do not. Implications of this finding for explaining observations of
astrophysical disks are discussed in \S~\ref{sec:discussion}.

\begin{figure}
  \begin{center}
    \begin{tabular}{c}
\resizebox{70mm}{!}{\includegraphics{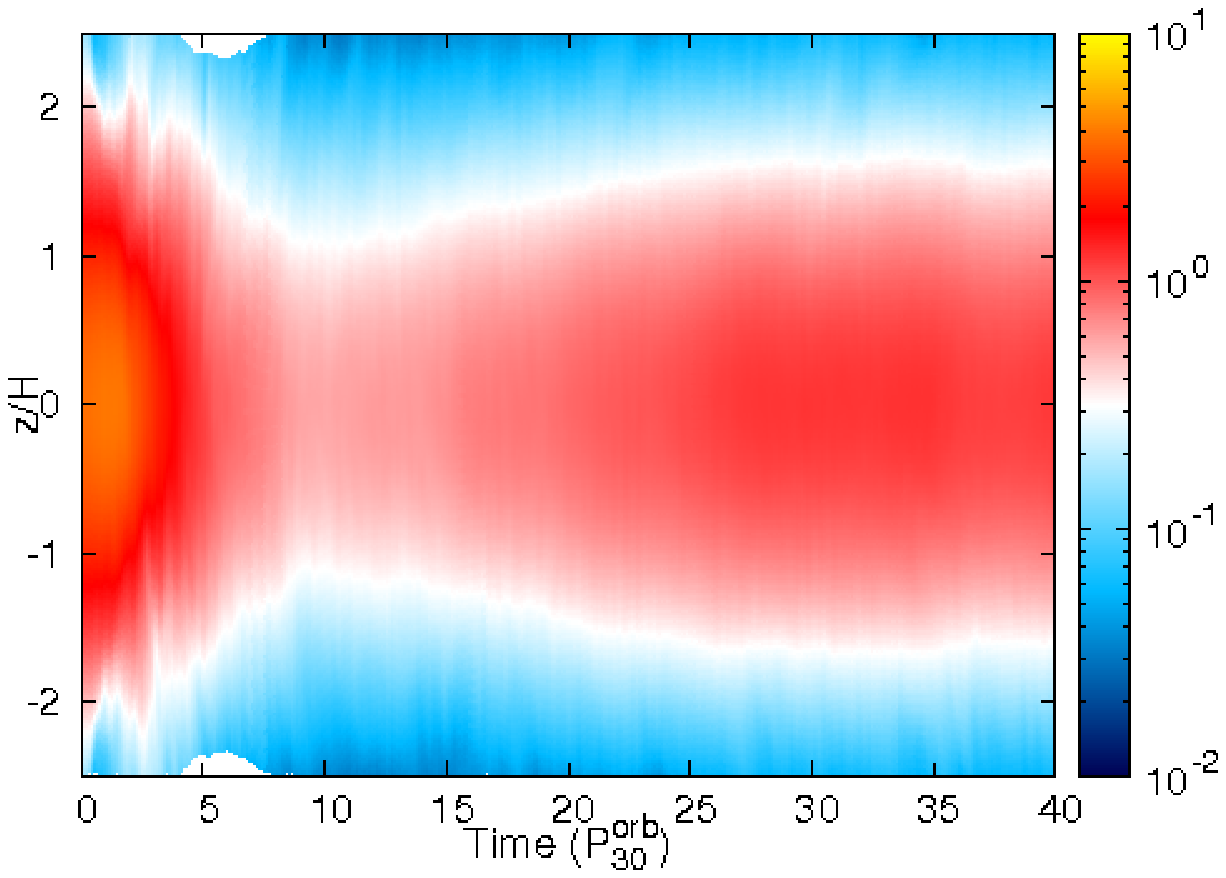}} \\
\resizebox{70mm}{!}{\includegraphics{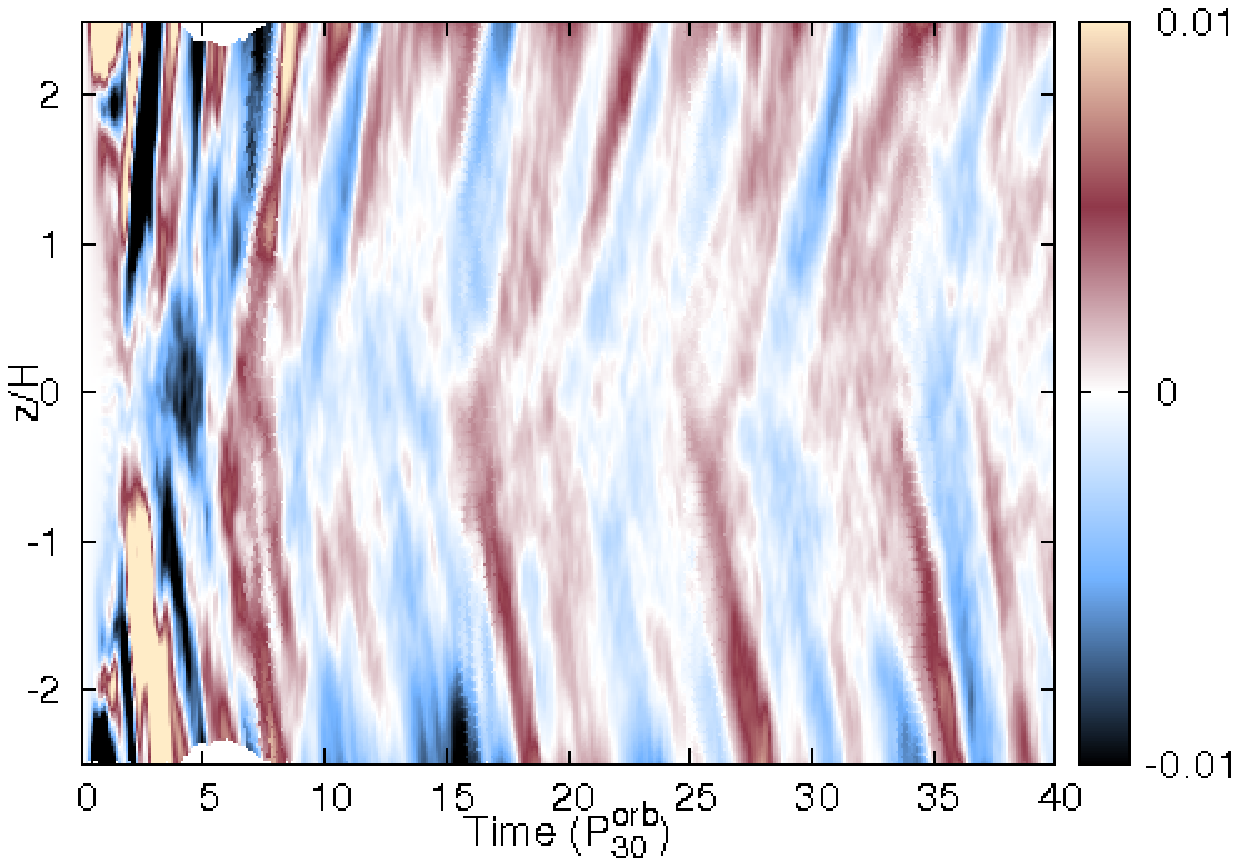}} \\
\resizebox{70mm}{!}{\includegraphics{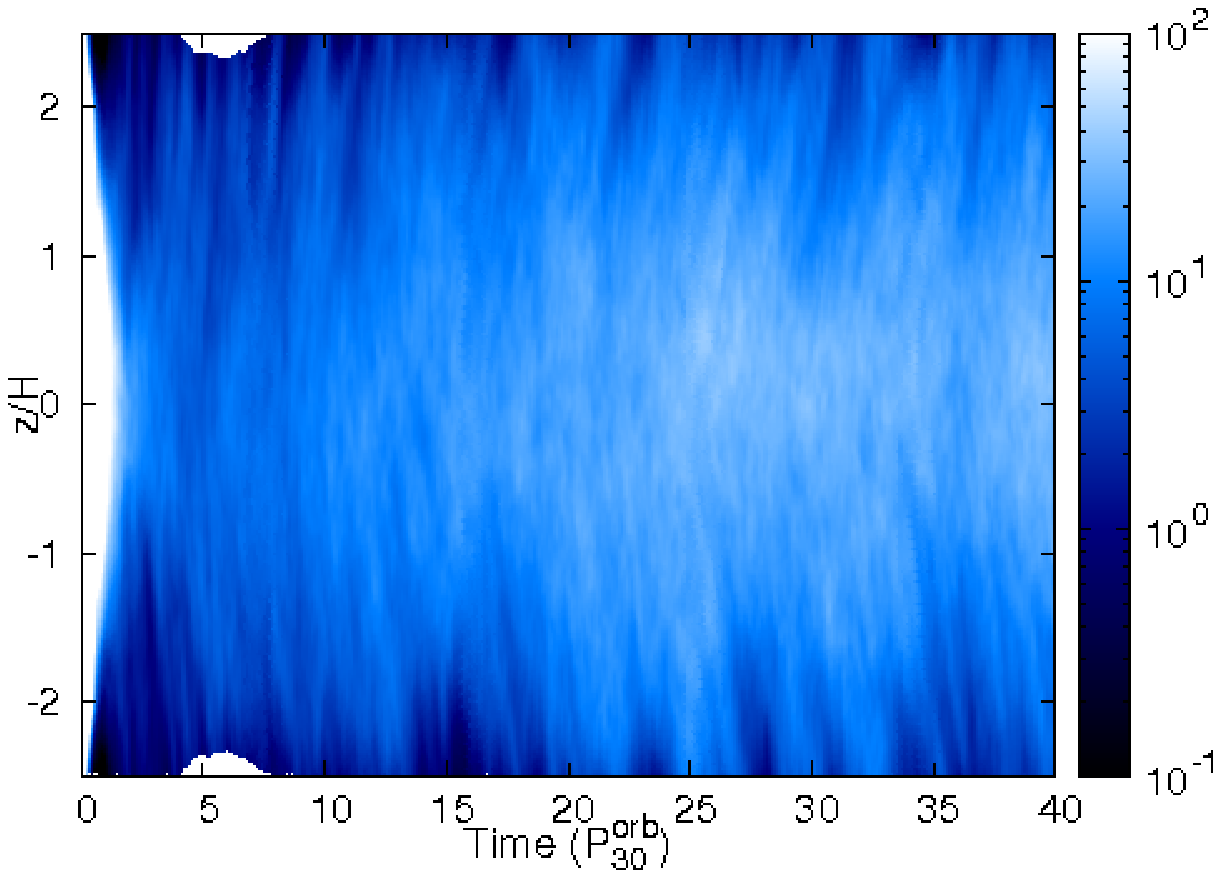}} \\
\resizebox{70mm}{!}{\includegraphics{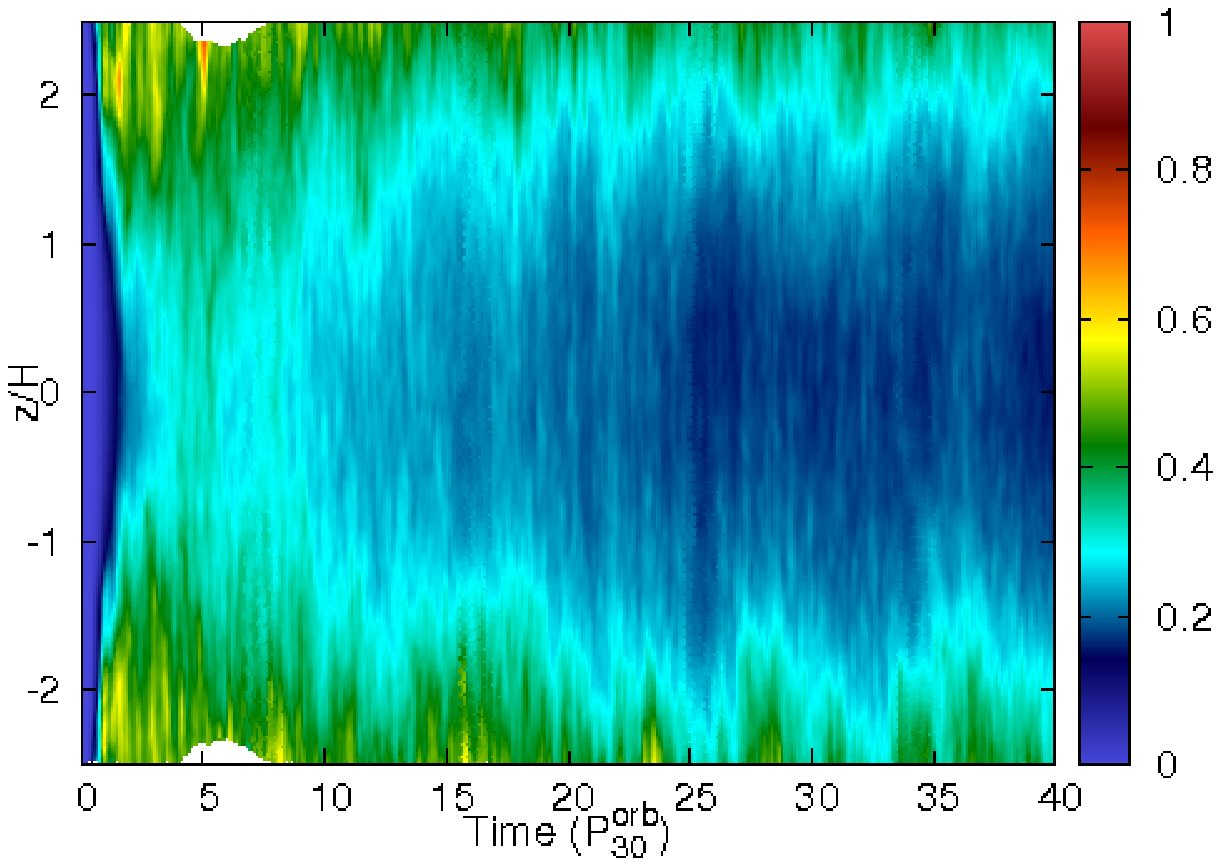}} \\
  \end{tabular}
   \caption{Time evolution of vertical profiles of various quantities
     in model VERT. Shown are: density (top), $B_{\phi}$ (upper
     middle), plasma-$\beta$ (lower middle), and the magnitude of the
     turbulent velocity normalised to the sound speed. To construct
     the profiles the simulation data was azimuthally averaged over
     the entire $\phi$-extent of the domain and radially between
     $15<r<25$. }
    \label{fig:vert_time}
  \end{center}
\end{figure}

\begin{figure*}
  \begin{center}
    \begin{tabular}{c}
\resizebox{170mm}{!}{\includegraphics{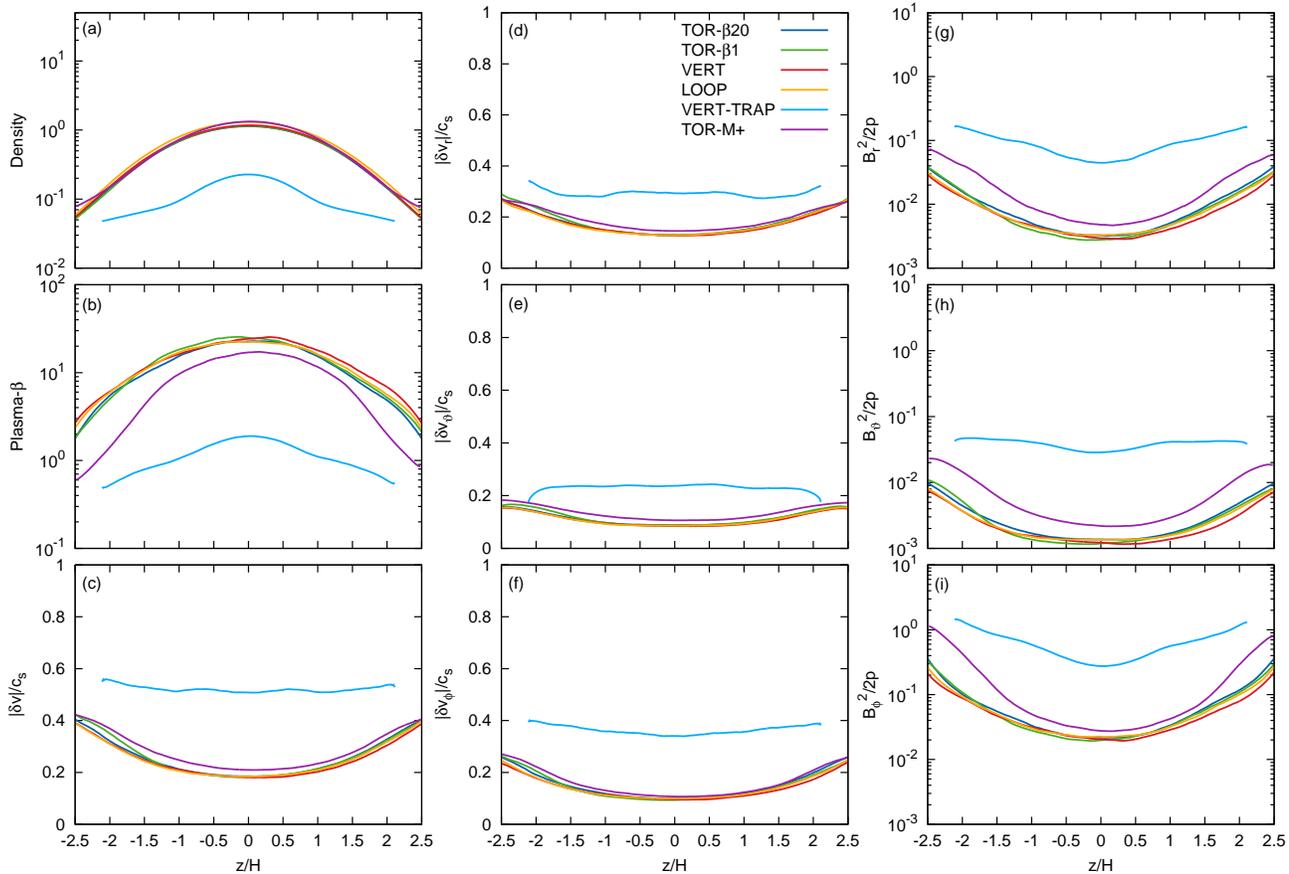}} \\
  \end{tabular}
   \caption{Vertical profiles of various quantities in the
     simulations. To construct the profiles the simulation data was
     azimuthally averaged over the entire $\phi$-extent of the domain,
     radially between $15<r<25$, and then time averaged over the time
     interval $20<t<40~P^{\rm orb}_{30}$ (except for model TOR-M+
     where the interval $25<t<40~P^{\rm orb}_{30}$ was used). Shown
     are: (a) density, (b) plasma-$\beta$ ($=2 p/B^2$), (c) turbulent
     gas velocity normalised to sound speed, (d)-(f) $r,\theta$, and
     $\phi$ turbulent velocity components normalised to the sound
     speed, and (g)-(i) $r,\theta$, and $\phi$ magnetic pressure
     normalised to gas pressure.}
    \label{fig:vert}
  \end{center}
\end{figure*}

Further similarities between the late-time evolution of the models are
found in the parameters $\langle \alpha_{\rm M}\rangle$ and $\langle
\alpha_{\rm K}\rangle $, which are defined to be the $r-\phi$
component of the Maxwell stress normalised by the magnetic pressure,
\begin{equation}
  \langle \alpha_{\rm M}\rangle = \frac{\langle -2
    B'_{r} B'_{\phi}\rangle }{\langle |B'|^2\rangle }, \label{eqn:alphaM}
\end{equation}
and the $r-\phi$ component of the Reynolds stress normalised by the turbulent
kinetic energy, 
\begin{equation}
  \langle \alpha_{\rm K}\rangle = \frac{\langle \rho v'_{r}
    v'_{\phi}\rangle }{\langle u_{\rm K'}\rangle }, \label{eqn:alphaK}
\end{equation}
where $\lb u_{\rm K'} \rb = \frac{1}{2} \rho |v'|^2$. The importance
of the parameter $\lb \alpha_{\rm K} \rb$ resides in turbulent kinetic
energy generation and the significant role of the correlation between
Reynolds stresses with mean flow shear as an energy injection term
\citep[][]{BH98, Kuncic:2004, Parkin:2014}. During the latter half of
the simulation $\langle \alpha_{\rm M}\rangle$ and $\langle
\alpha_{\rm K}\rangle $ reach quasi-steady values of $\sim 0.47$ and
$\sim 0.28$, respectively, with little difference between models
TOR-$\beta1$, TOR-$\beta20$, VERT, and LOOP
(Table~\ref{tab:models}). The values of $\lb \alpha_{\rm K}\rb$ from
the models are consistent with, but slightly higher than, the value of
0.25 reported for stratified shearing-box models by
\cite{Brandenburg:1995}. We note that $\langle \alpha_{\rm M}\rangle$
reaches a steady value faster than $\lb \alpha_{\rm P} \rb$, allowing
safe comparisons to be made against previous shorter run time global
models. As such, we find good agreement for time averaged values of
$\langle \alpha_{\rm M}\rangle$ with the studies by \cite{Hawley:2011,
  Hawley:2013}, \cite{Beckwith:2011}, and \cite{Parkin:2013,
  Parkin:2013b}. The ratio of Maxwell-to-Reynolds stress, $-\langle
B'_{r}B'_{\phi}\rangle / \lb \rho v'_{r}v'_{\phi} \rb = 2.55-2.69$ is
also consistent with previous global disk studies \citep{Fromang:2006,
  Beckwith:2011, Parkin:2014}.

\subsubsection{Disk structure}
\label{subsubsec:disk_structure}

Following the development of the MRI in the simulations, the disk
structure in models TOR-$\beta1$, TOR-$\beta20$, VERT, and LOOP
evolves towards a quasi-steady state. The time evolution of vertical
profiles of the gas density, toroidal magnetic field, plasma-$\beta$,
and turbulent gas velocity are shown for model VERT in
Fig.~\ref{fig:vert_time}. During the initial transient phase at the
start of the simulation the rapid growth of accretion stresses drives
mass out of the disk. At later times the density distribution in the
disk reaches a close equilibrium between mass inflow/outflow. The
heightened level of turbulent activity during the transient phase is
evident from the turbulent gas velocity, which rises at the start of
the run and then falls at later times. There is a coincident rise and
fall in magnetic field strength, illustrated by the plasma-$\beta$
parameter. Quasi-periodic oscillations indicative of dynamo cycles are
apparent in the toroidal magnetic field (upper middle panel in
Fig.~\ref{fig:vert_time}). The oscillation period of $4\;P^{\rm
  orb}_{30}$ is in good agreement with that observed in previous
shearing-box \citep[e.g.][]{Davis:2010, Gressel:2010, Shi:2010,
  Simon:2012} and global models \citep{Beckwith:2011, Flock:2012,
 Parkin:2014, Suzuki:2014}. Interestingly, much shorter period
fluctuations can be seen in the plasma-$\beta$ and turbulent velocity,
which may be associated with the intermittency of the turbulence, or a
secondary cycle.

Examining the vertical structure of the simulated disks (time-averaged
over the latter half of the run) one sees markedly similar profiles
for model TOR-$\beta$1, TOR-$\beta$20, VERT, and LOOP. The gas density
(Fig.~\ref{fig:vert}(a)) follows a roughly Gaussian shape, consistent
with previous global disk studies \citep[e.g.][]{Flock:2011,
  Suzuki:2014}. The magnetic field strength is sub-thermal close to
the mid plane, illustrated by $\beta \gg 1$ in Fig~\ref{fig:vert}(b)
at $z\simeq 0$, and rises ($\beta \rightarrow1$) with increasing
height in agreement with \cite{Fromang:2006}, \cite{Flock:2011},
\cite{Beckwith:2011}, and \cite{Parkin:2014}. Similarly, the turbulent
gas velocities are lowest close to the mid plane, being subsonic at
all heights, and increasing as one tends towards the upper/lower grid
boundary. Inspecting the separate components of the turbulent gas
velocity, shown in Fig~\ref{fig:vert}(d)-(f), the radial and azimuthal
fluctuations are larger than those in the vertical direction.  Similar
behaviour for turbulent gas velocities was observed by
\cite{Fromang:2006, Flock:2011, Beckwith:2011}, and
\cite{Parkin:2014}, who found $|\delta v|/c_{\rm s} \simeq 0.1$ at the
mid plane for the different velocity components.

\begin{figure}
  \begin{center}
    \begin{tabular}{c}
\resizebox{80mm}{!}{\includegraphics{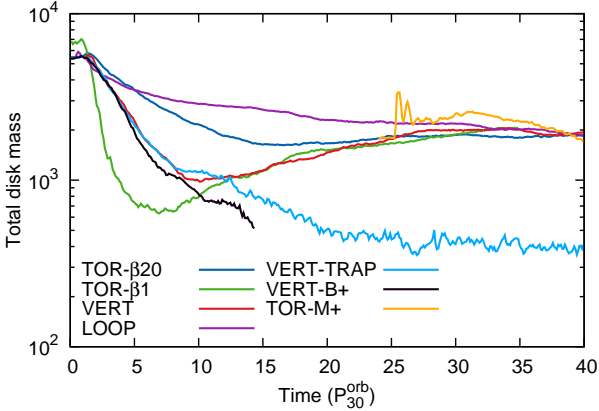}} \\
  \end{tabular}
   \caption{The evolution of the total disk mass. Corresponding
     time-averaged values can be found in Table~\ref{tab:models}.}
    \label{fig:total_mass}
  \end{center}
\end{figure}

\begin{figure*}
  \begin{center}
    \begin{tabular}{c}
\resizebox{170mm}{!}{\includegraphics{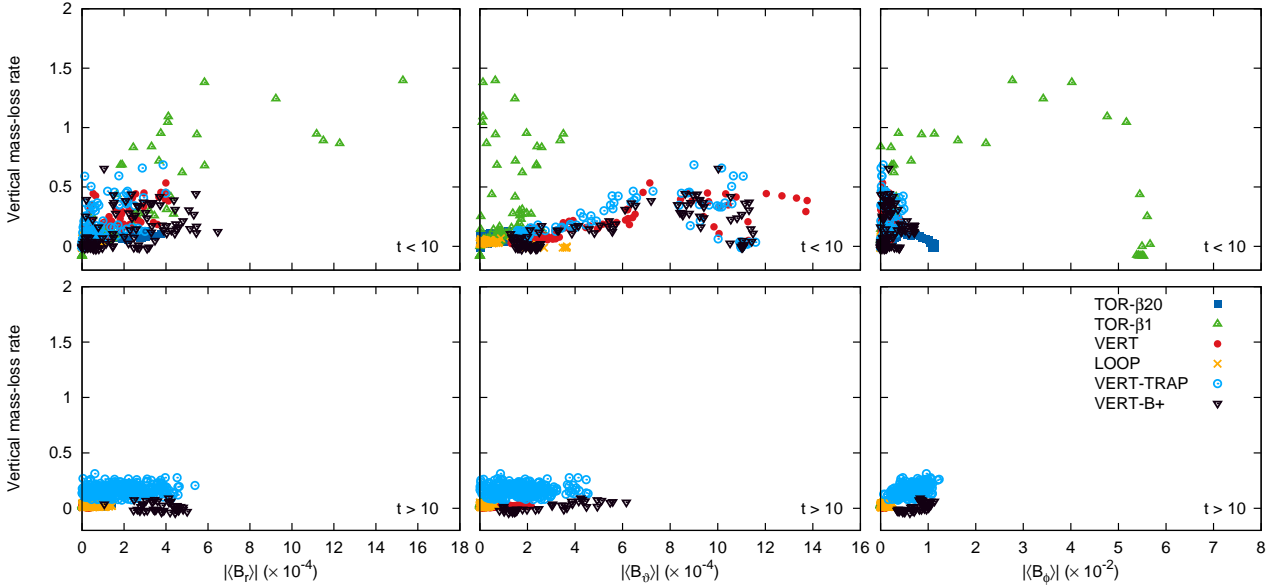}} \\
  \end{tabular}
   \caption{Correlation between the vertical mass-loss rate,
     $\dot{M}_{\rm vert}$, and the magnitude of the volume averaged
     magnetic field. The columns (from left to right) show the
     correlation with: $|\lb B_{r} \rb|$, $| \lb B_{\theta} \rb|$, and
     $|\lb B_{\phi} \rb|$. The upper and lower rows show data points
     sampled every $0.1 \ts P^{\rm orb}_{30}$ from the time intervals
     $t<10 \ts P_{30}^{\rm orb}$ and $t>10 \ts P_{30}^{\rm orb}$,
     respectively. }
    \label{fig:B_mdotvert}
  \end{center}
\end{figure*}

\subsubsection{Mass-loss}

The establishment of turbulence and associated angular momentum
transport in the simulations leads to evolution of the total disk
mass, as illustrated by Fig.~\ref{fig:total_mass}. All of the models
show a decrease in the total disk mass during the transient evolution
phase, followed by a flattening as the turbulent activity subsides and
a balance is reached between rates of mass leaving and entering the
disk via accretion and mass injection, respectively. The total disk
mass during the quasi-steady state, $M_{\rm QSS}$, is noted in
Table~\ref{tab:models}. Reasonable agreement is seen in $M_{\rm QSS}$
between models TOR-$\beta1$, TOR-$\beta20$, VERT, and LOOP, despite
TOR-$\beta1$ having a slightly higher initial disk mass of $M_{0}=
6759$ (compared to $5376-5477$ for the other models). On average, the
total disk mass during the quasi-steady state is $33\%$ of the initial
disk mass.

Surface integrated mass fluxes have been computed to assess the rates
at which mass is leaving the disk in the radial, $\dot{M}_{\rm rad}$,
and vertical, $\dot{M}_{\rm vert}$, directions. It is clear from
Table~\ref{tab:models} that the radial mass flux dominates over the
vertical mass flux for models TOR-$\beta1$, TOR-$\beta20$, VERT, and
LOOP.

Models of stratified disks with a net vertical magnetic field, in both
the shearing-box\footnote{It is noteworthy that the mean radial and
  azimuthal magnetic field can evolve in a stratified shearing-box
  whereas the mean vertical field cannot \citep[see
  \S~\ref{sec:mean_field} and][for further
  discussion]{Parkin:2013b}. Therefore, it is the magnitude of the
  vertical component of the initial magnetic field, and not the field
  topology that is critical for shearing-box models.} and global
setting, suggest a strong correspondence between the vertical magnetic
field strength and the rate of vertical mass-loss \citep{Suzuki:2009,
  Suzuki:2014, Bai:2013}. To examine whether a similar trend is
observed in the global disk models presented in this work, in
Fig.~\ref{fig:B_mdotvert} we plot data-points for $\dot{M}_{\rm vert}$
and $|\lb B \rb|$. For all models except TOR-$\beta$1 there is a
suggestion that larger $\dot{M}_{\rm vert}$ values are coincident with
a stronger vertical magnetic field strength ($|\lb B_{\theta}
\rb|$). This is most clearly seen at early times in the simulations
(top row of Fig.~\ref{fig:B_mdotvert}). However, a contrasting trend
is seen for model TOR-$\beta$1, where the largest vertical mass-loss
rates arise during the transient phase of the simulation when $|\lb
B_{\theta} \rb|$ is relatively small. Interestingly, the large
$\dot{M}_{\rm vert}$ at early times ($t < 10\;P^{\rm orb}_{30}$) in
TOR-$\beta$1 correlates well with radial magnetic field ($|\lb B_{r}
\rb|$), with a weaker correlation with the azimuthal magnetic field
($|\lb B_{\phi} \rb|$). The data points for TOR-$\beta$1 in the top
row of Fig.~\ref{fig:B_mdotvert} are consistent with vertical
mass-loss resulting from stresses and magnetic pressure gradients
associated with the disruption of the strong toroidal field during the
early-time transient phase.

Before closing this section it is pertinent to recall an important
finding from the recent study by \cite{Fromang:2013}, namely that the
vertical mass-loss rate depends on the vertical extent of the
simulation domain. This behaviour is linked to the position of the
critical points of the wind launching, and whether the adopted
vertical grid extent captures all of the relevant critical points. Due
to the large computational expense of performing global disk
simulations at the resolution of the models presented in this work, we
must defer a comprehensive study of similar models with considerably
larger vertical extent to a future study. As such, we urge that the
differences in vertical mass-loss rate between the models in this work
should be taken as illustrative (because the flow velocities are
subsonic as the gas exits the vertical grid boundary). Further work
will be needed to establish the exact quantitative details of magnetic
flux transport in thin accretion disks possessing a large-scale
vertical field, and whether inward or outward radial flux transport
occurs \citep[see, e.g.,][]{Lubow:1994, Beckwith:2009, Guilet:2012,
  McKinney:2012}.

\subsubsection{Resolution of the MRI}

To demonstrate that the simulations are sufficiently well resolved,
and that their common late-time evolution is not an artefact of
under-resolved MRI modes, we calculate a resolvability factor, $N_{\rm
  i}$, which is defined to be the fraction of cells in the disk body
that resolve the wavelength of the fastest growing MRI mode,
$\lambda_{\rm MRI-i}$, with at least 8 cells\footnote{This equates to
  measuring the fraction of cells which have a ``quality factor''
  \citep{Noble:2010, Hawley:2011} which is 8 or better throughout the
  disk - see also \cite{Sorathia:2012} and
  \cite{Parkin:2013b}.}. Defining,
\begin{equation}
  \lambda_{\rm MRI-i} = \frac{2 \pi |v_{\rm A i}| r \sin
    \theta}{v_{\phi}}, \label{eqn:lambdaMRI}
\end{equation}
where $i=r,\theta,\phi$, and $v_{\rm A i}=B_{\rm i}/\sqrt{\rho}$ is
the Alfv{\' e}n speed. Time-averaged values for the resolvability
factors (computed for the quasi-steady turbulent state) are noted in
Table~\ref{tab:models}. The values are similar between the models,
consistent with their broad similarity during the latter half of the
runs. Moreover, the values are indicative of well-resolved MRI-driven
turbulence \citep{Sorathia:2012, Parkin:2013b}.

\begin{figure}
  \begin{center}
    \begin{tabular}{c}
\resizebox{75mm}{!}{\includegraphics{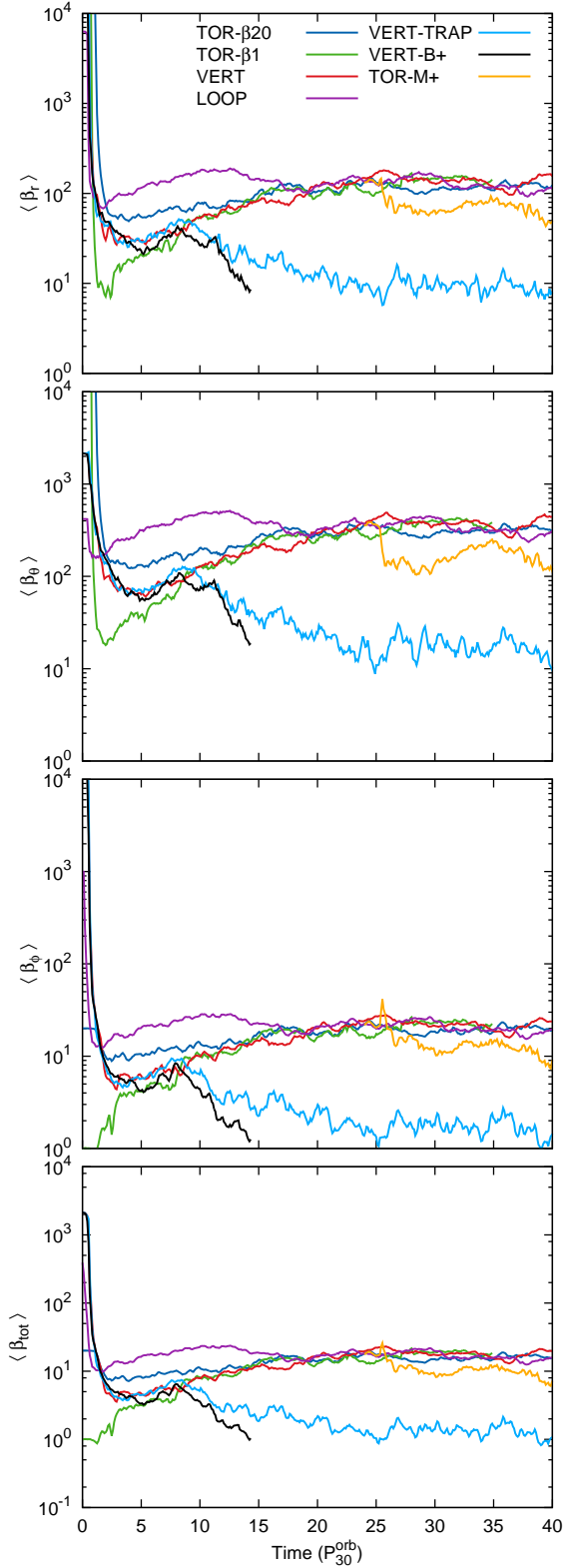}} \\
  \end{tabular}
   \caption{The volume-averaged (over the region $15<r<25, |z| < 2H$)
     plasma-$\beta$ for models TOR-$\beta20$, TOR-$\beta1$, VERT,
     LOOP, VERT-TRAP, and TOR-M+. Respective plasma-$\beta$ values
     for different magnetic field components are shown (from top to
     bottom): $\lb \beta_r \rb$, $\lb \beta_{\theta} \rb$, $\lb
     \beta_{\phi} \rb$, and $\lb \beta_{tot} \rb$ (i.e. all
     components). Corresponding time-averaged values can be found in
     Table~\ref{tab:models}.}
    \label{fig:beta}
  \end{center}
\end{figure}

\begin{figure}
  \begin{center}
    \begin{tabular}{c}
\resizebox{75mm}{!}{\includegraphics{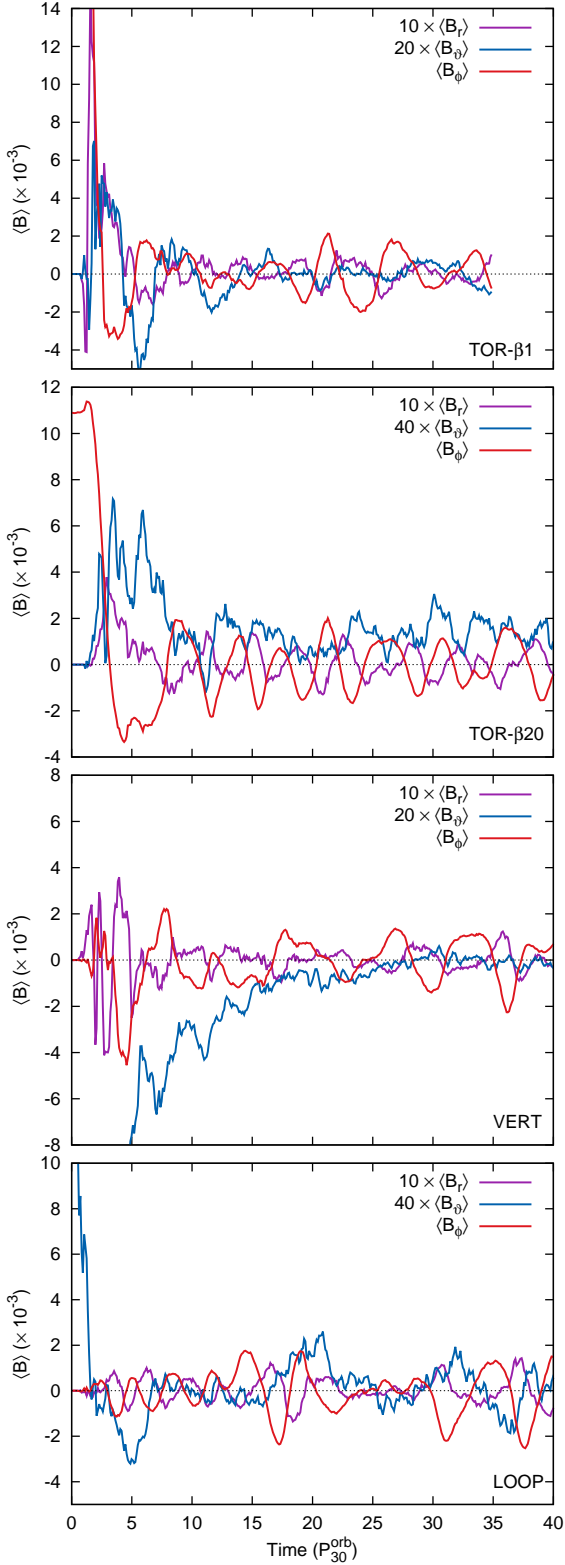}} \\
   \end{tabular}
   \caption{Mean magnetic field components for models TOR-$\beta1$
     (top), TOR-$\beta20$ (upper middle), VERT (lower middle), and
     LOOP (bottom). Values correspond to volume-averages over the
     region $15<r<25, |z| < 2H$. To aid comparison with $\lb B_{\phi}
     \rb$, the curves for $\lb B_{r} \rb$ and $\lb B_{\theta} \rb$
     have been multiplied by constant factors (noted in the plots).}
    \label{fig:Bav}
  \end{center}
\end{figure}

\subsubsection{Magnetic field evolution}
\label{subsec:mag_fields}

The behaviour of the magnetic energies in the different models is in
keeping with their respective initial field topologies
(Fig.~\ref{fig:beta}). For instance, defining the plasma-$\beta$ as
$\beta_{i}=|B_{i}|^{2}/2p$, where $i=r,\theta,\phi$ or ``tot''
(i.e. all components), models TOR-$\beta1$ and TOR-$\beta20$ are
clearly dominated by toroidal magnetic field during the first few
orbits, whereas model VERT (which started with a purely vertical
field) has a smaller $\lb \beta_{\theta} \rb$ initially. In fact,
after a few orbits, the conversion of poloidal magnetic field into
toroidal field by the disk rotation results in $\lb \beta_{\phi} \rb$
clearly dominating $\lb \beta_{\rm tot} \rb$ in all of the
models. Consistent with the diagnostics discussed in the previous
sections, during the quasi-steady state ($20 < t < 40 \ts P_{30}^{\rm
  orb}$), the evolution of $\beta$ in models TOR-$\beta1$,
TOR-$\beta20$, VERT, and LOOP closely matches one
another. Time-averaged values are also in good agreement
(Table~\ref{tab:models}). Curiously, the vertical magnetic field
strength is weakest for model VERT, despite it possessing the
strongest vertical field in the initial conditions. However, the
indication from the plot of $\lb \alpha_{\rm P} \rb$ in
Fig.~\ref{fig:alpha} is that the transient phase lasts the longest for
this model. Hence, a prolonged phase with a higher accretion rate will
drain the poloidal magnetic fields from the disk. The evolution of the
mean magnetic fields, discussed below, supports this explanation.

Focusing next on the evolution of the mean (volume-averaged) magnetic
fields shown in Fig.~\ref{fig:Bav}, one sees both similarities and
marked differences between the models. Firstly, they show different
evolution for the mean vertical field, $\lb B_{\theta} \rb$. The
initial vertical field decays throughout the simulation for model
VERT, as anticipated from the discussion of mean field evolution in
\S~\ref{sec:mean_field}. Second, all models show anti-correlated,
quasi-periodic oscillations in $\lb B_{r} \rb$ and $\lb B_{\phi}
\rb$. However, these oscillations are most evident for model
TOR-$\beta20$, with a weaker indication for VERT and LOOP. Thirdly,
there is a suggestion of quasi-periodic oscillations in $\lb
B_{\theta} \rb$ for models TOR-$\beta1$, TOR-$\beta20$, and LOOP,
although of quite different period, being more erratic for
TOR-$\beta20$ whilst more slowly varying for LOOP. Hence, although
diagnostics of the turbulence (such as $\lb \alpha_{\rm P} \rb$ and
$\beta$) do not display considerable differences between the models
during their late-time evolution, the mean magnetic fields do. The
intriguing implication is that the large-scale turbulent dynamo
retains some memory of the initial conditions for a longer time than
the turbulent stresses, and may exhibit different behaviour (on very
long timescales) as a consequence. With regard to characterising the
dynamo in the disk, it is noteworthy that the time-averaged radial and
azimuthal fields are close to zero (Table~\ref{tab:models}), whereas
the vertical field maintains a (weak) net value.

\subsubsection{Do stresses correlate with vertical field?}
\label{subsec:alphaP_Btheta}

Local (shearing-box) studies of magnetorotational turbulence indicate
a correlation between the vertical field strength and the turbulent
stresses \citep{Hawley:1995, Sano:2004}. This correlation has also
been cast as a relation between the wavelength of the fastest growing
vertical MRI mode and the stress \citep{Pessah:2007, Sorathia:2010,
  Sorathia:2012, Beckwith:2011}. Considering the ab initio vertical
flux conservation inherent to the shearing-box \citep[see
\S~\ref{sec:mean_field} and][]{Hawley:1995}, it is pertinent to
investigate this relation using the global models in this work.

\begin{figure*}
  \begin{center}
    \begin{tabular}{c}
\resizebox{170mm}{!}{\includegraphics{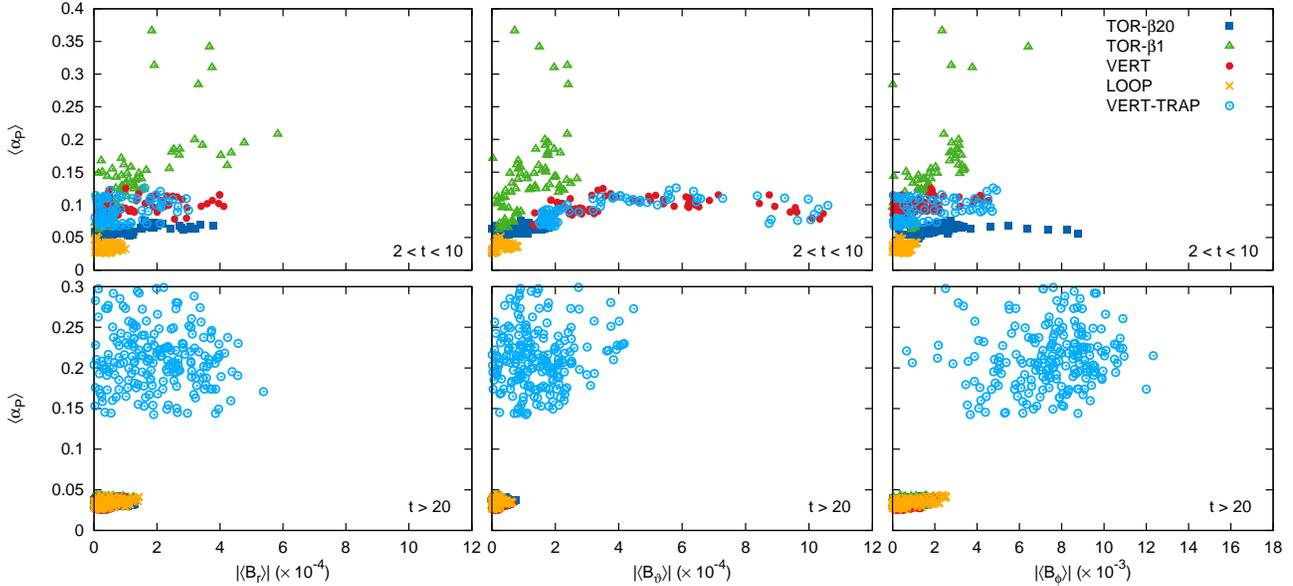}} \\
  \end{tabular}
   \caption{Correlation between the pressure-normalised stress
     parameter, $\lb \alpha_{\rm P} \rb$, and the magnitude of the
     volume averaged magnetic field. The columns (from left to right)
     show the correlation with: $|\lb B_{r} \rb|$, $| \lb B_{\theta}
     \rb|$, and $|\lb B_{\phi} \rb|$. The upper and lower rows show
     data points sampled every $0.1 \ts P^{\rm orb}_{30}$ from the
     time intervals $2<t<10 \ts P_{30}^{\rm orb}$ and $t>20 \ts
     P_{30}^{\rm orb}$, respectively. }
    \label{fig:alphaP_Btheta}
  \end{center}
\end{figure*}

To search for possible correlations between turbulent stresses and
mean magnetic field components, in Fig.~\ref{fig:alphaP_Btheta} we
plot the data-points for $\lb \alpha_{\rm P} \rb$ against those for
$|\lb B_{r} \rb|$ (left column), $| \lb B_{\theta} \rb|$ (middle
column), and $|\lb B_{\phi} \rb|$ (right column). The points are
sampled every $0.1 \ts P_{30}^{\rm orb}$ in time through the
simulation, such that there is a relation between the clustering of
points and the duration of time that a specific value was possessed by
the model. The upper panel of Fig.~\ref{fig:alphaP_Btheta} shows
points for the time interval $2<t<10 \ts P_{30}^{\rm orb}$, which
corresponds to the transient episode close to the start of the
simulations. During this phase the models are in the process of
expelling, or turbulently corrupting, the initial field
configuration. As such, the vertical field initially present in models
VERT and LOOP has not been entirely removed and a general trend of
higher $\lb \alpha_{\rm P} \rb$ with higher $|\lb B_{\theta} \rb |$
can be seen. However, model TOR-$\beta1$ achieves the highest stresses
during its transient phase ($\lb \alpha_{\rm P} \rb \simeq 0.37$),
with a less apparent correlation between $\lb \alpha_{\rm P} \rb $ and
any of $|\lb B_{r} \rb|$, $|\lb B_{\theta} \rb|$, or $|\lb B_{\phi}
\rb|$.

By the time the quasi-steady state is reached, models TOR-$\beta1$,
TOR-$\beta20$, VERT, and LOOP have removed any prominent vertical
field, showing similar vertical magnetic field strengths
(\S~\ref{subsec:mag_fields}). At these late-times ($t>20 \ts
P_{30}^{\rm orb}$) the data-points are heavily clustered around $\lb
\alpha_{\rm P} \rb \simeq 0.025-0.045$ and $\lb B_{\theta} \rb \sim
0$.

In summary, during the transient phase close to the beginning of the
runs the models suggest a correlation between $\lb \alpha_{\rm P} \rb$
and $\lb B_{\theta} \rb$. However, at later times there is little sign
of such a trend, mainly due to the effective removal of a mean
vertical field by the evolution of the disk. Hence, it seems less a
question of whether turbulent stresses, and thus the efficiency of
accretion, correlate with vertical magnetic field strength, and more
one of whether a global turbulent disk retains a vertical field in the
face of said accretion. 

\subsection{Trapping magnetic field (model VERT-TRAP)}
\label{subsec:vert-vbc}

The results of the previous section highlight the common late-time
evolution for models that are initialised with different magnetic
field configurations and for which the radial velocity at the inner
radial edge of the disk (the simulation boundary in this case) is
unconstrained. There are, however, many examples of astrophysical
scenarios in which the radial velocity may be limited by large scale
effects. For instance, clumping due to self-gravity in an extended
disk \citep{Shlosman:1987, Goodman:2003, Rafikov:2009,
  Hopkins:2013}. Also, the excessive build up of magnetic flux close
to a black hole leading to a magnetically arrested disk
\citep{Narayan:2003, Igumenshchev:2003, Igumenshchev:2008}. In this
section we examine the impact of restricting the outflow of material
at the inner radial disk edge using model VERT-TRAP (which adopts the
viscous-outflow boundary condition described in
\S~\ref{subsec:hydromodel}).

\begin{figure}
  \begin{center}
    \begin{tabular}{c}
\resizebox{80mm}{!}{\includegraphics{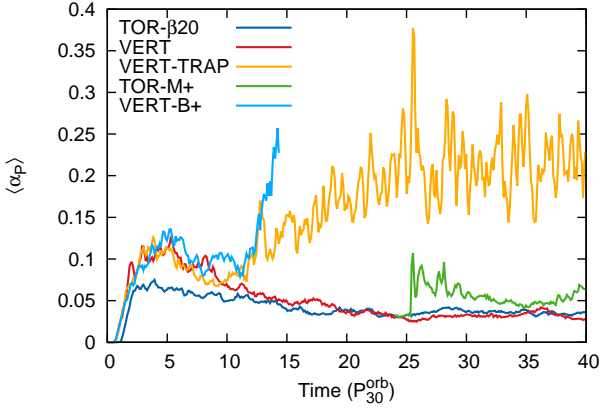}} \\
   \end{tabular}
   \caption{The time evolution of $\langle\alpha_{\rm P}\rangle$
     computed for the disk body ($15<r<25$, $|z|<2H$) for models
     TOR-$\beta$20, VERT, VERT-TRAP (\S~\ref{subsec:vert-vbc}),
     VERT-B+ (\S~\ref{subsec:vert-Bthadd}), and TOR-M+
     (\S~\ref{subsec:tor-inj}).}
    \label{fig:alpha_tests}
  \end{center}
\end{figure}

The evolution of $\lb \alpha_{\rm P} \rb$ for model VERT-TRAP is shown
in Fig.~\ref{fig:alpha_tests}. Over the first $\sim 10 \ts P^{\rm
  orb}_{30}$ of the run the turbulence develops in a very similar
manner to model VERT; both models display transient accretion
associated with the initiation of magnetorotational turbulence in the
disk, which saturates at $t\simeq 5 \ts P^{\rm orb}_{30}$ and
gradually declines thereafter. However, in model VERT-TRAP the
accretion flow builds-up at the inner radial disk
edge\footnote{\cite{Suzuki:2014} adopted a similar radial boundary
  condition at the inner grid edge but did not constrain the mass-flux
  to the extent that mass would build-up. Instead, they tuned their
  boundary condition to prevent excessive mass-loss during the initial
  transient whilst allowing a flow of mass consistent with the
  accretion rate induced by the turbulence during the quasi-steady
  phase.}, leading to a compression of magnetic field. Consequently,
the magnetisation of the disk rises and with it the turbulent stress
($\lb \alpha_{\rm P} \rb$). During the latter half of the simulation
($t>20 \ts P^{\rm orb}_{30}$) $\lb \alpha_{\rm P} \rb $ and
$\beta_{\rm tot }$ reach time-averaged values of 0.21 and 1.4,
respectively (Table~\ref{tab:models}). Large amplitude fluctuations
for $\lb \alpha_{\rm P} \rb $ bear similarities with those observed in
strongly magnetized ($\beta \sim 1- 5$) disk simulations by
\cite{Machida:2000}. The vertical field becomes an order of magnitude
more magnetized for VERT-TRAP than for VERT, with $\beta_{\theta}
\simeq 10$ and 139 for the respective models. Strong magnetisation in
the disk leads to a ratio of Maxwell-to-Reynolds stress, $-\langle
B'_{r}B'_{\phi}\rangle / \lb \rho v'_{r}v'_{\phi} \rb = 5.79$. Hence,
$\lb \alpha_{\rm P} \rb$ is considerably dominated by magnetic
stresses in the field-trapped state.

\begin{figure}
  \begin{center}
    \begin{tabular}{c}
\resizebox{70mm}{!}{\includegraphics{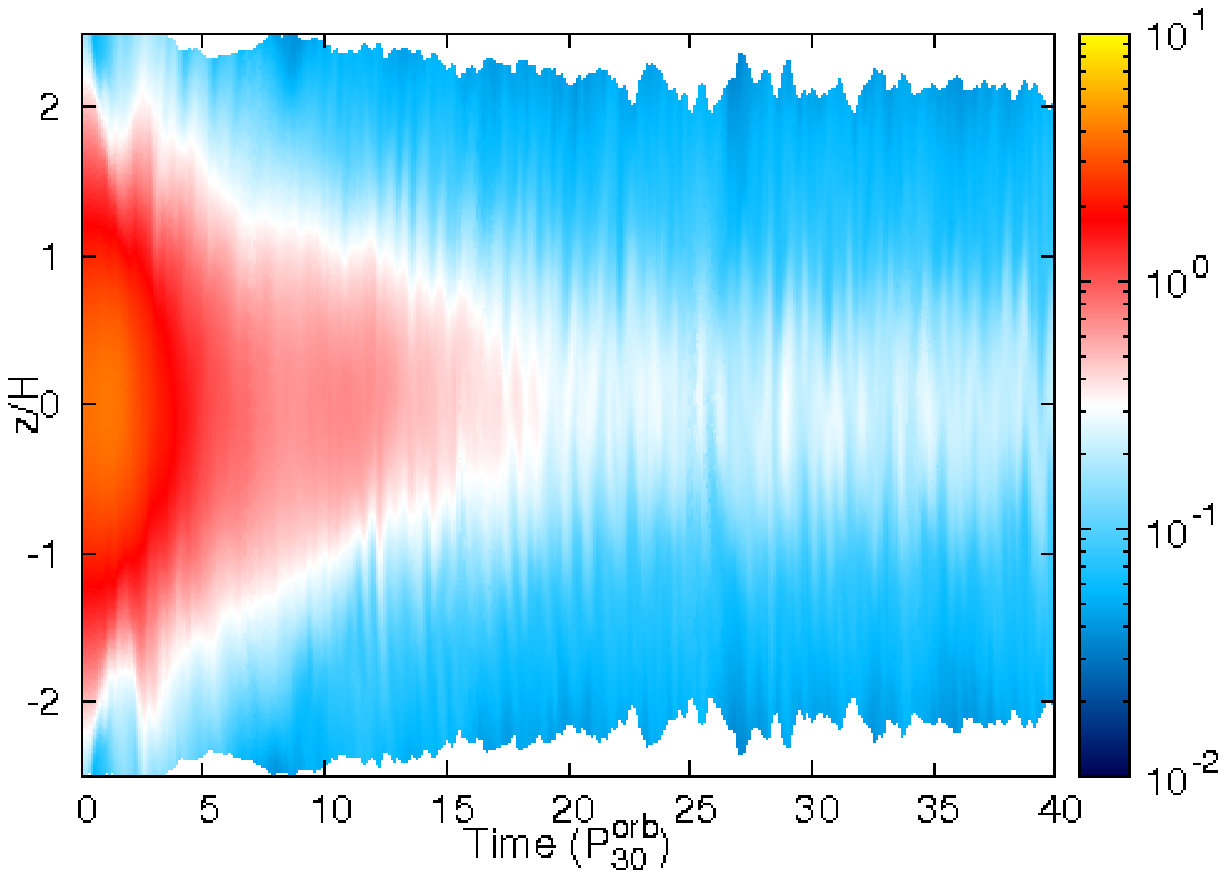}} \\
\resizebox{70mm}{!}{\includegraphics{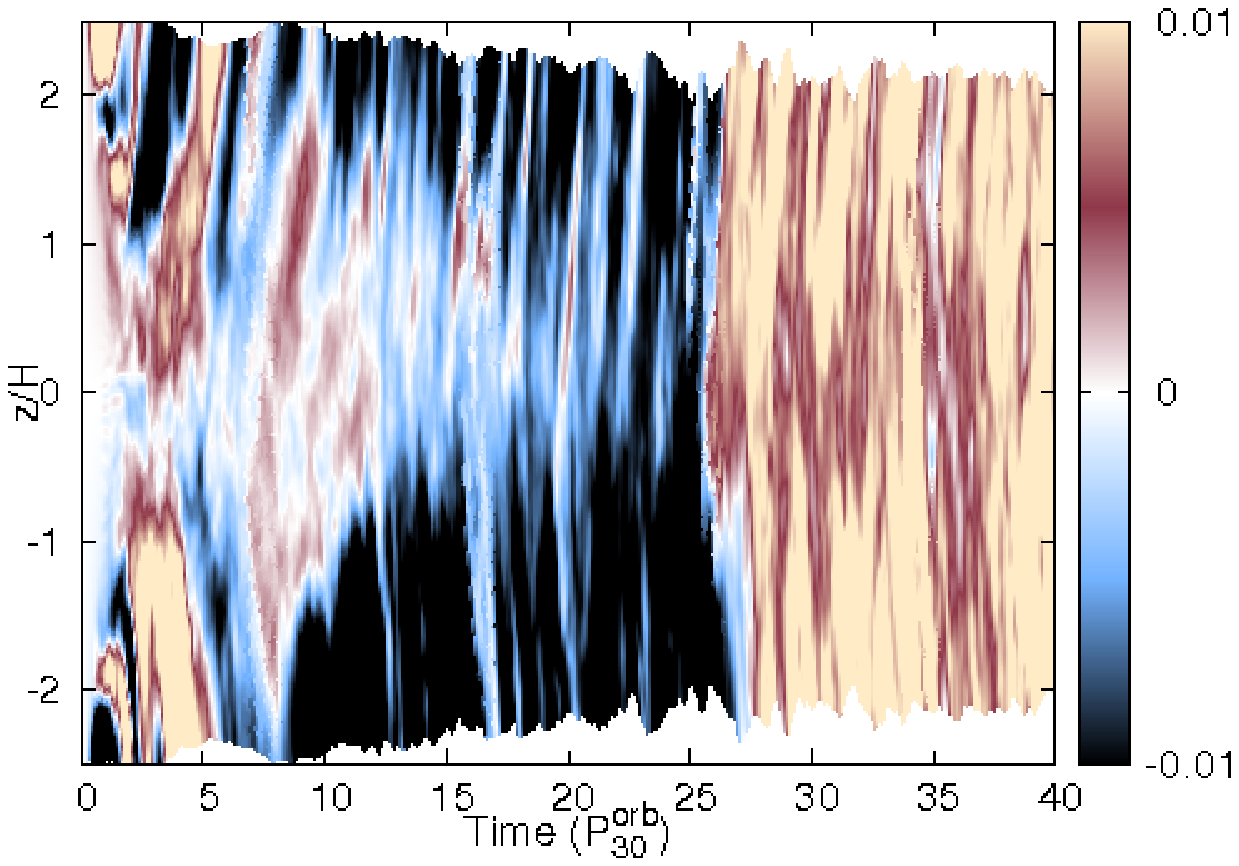}} \\
\resizebox{70mm}{!}{\includegraphics{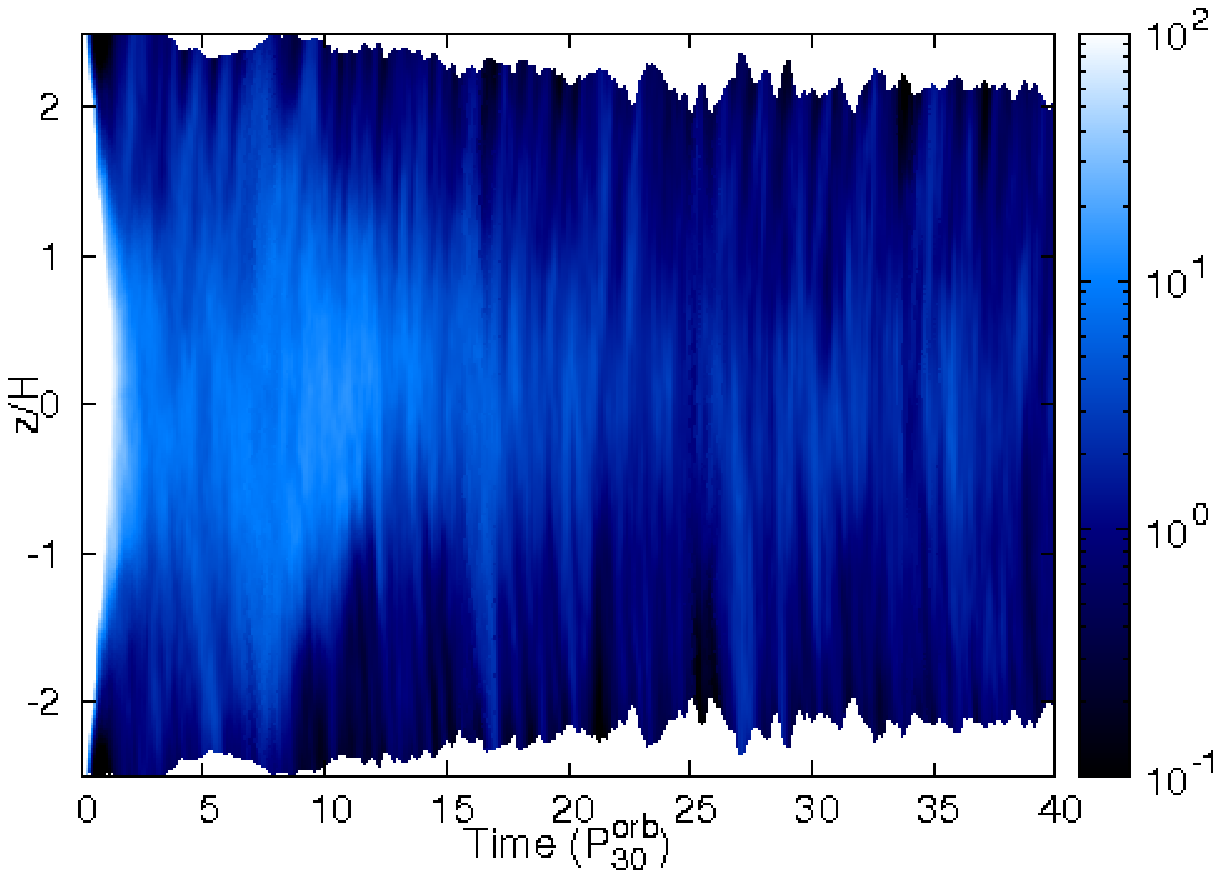}} \\
\resizebox{70mm}{!}{\includegraphics{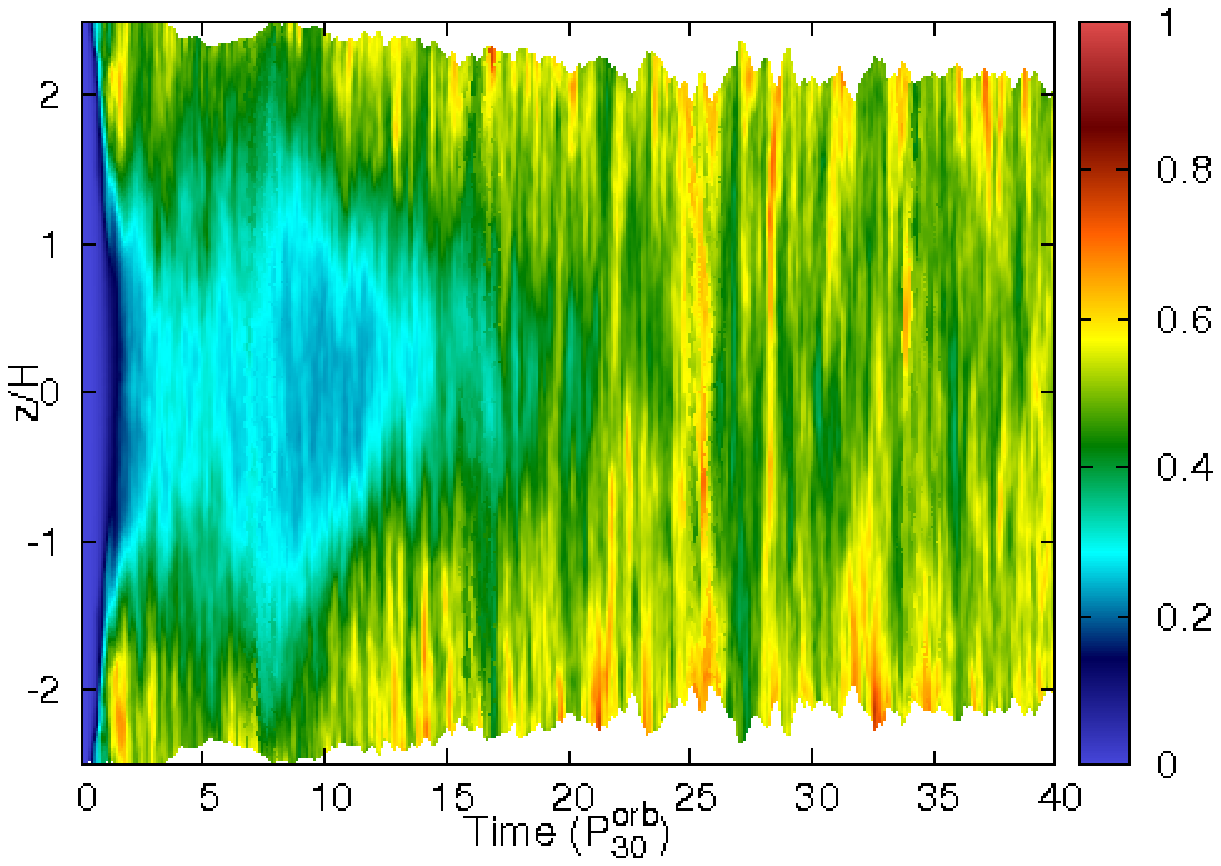}} \\
   \end{tabular}
   \caption{Same as Fig.~\ref{fig:vert_time} except for model VERT-TRAP}
    \label{fig:vert_time_vt}
  \end{center}
\end{figure}

\begin{figure}
  \begin{center}
    \begin{tabular}{c}
\resizebox{80mm}{!}{\includegraphics{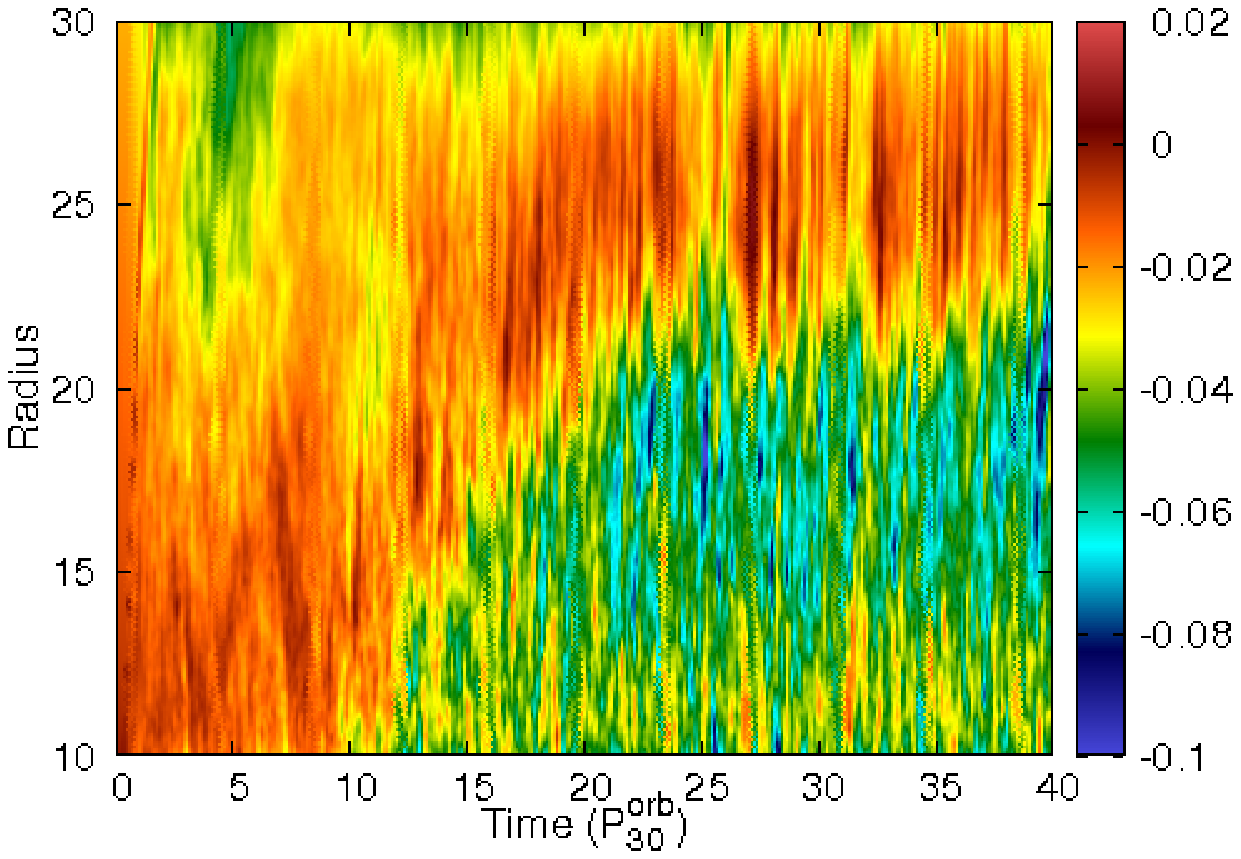}} \\
   \end{tabular}
   \caption{Time evolution of the fractional variation of the
     rotational velocity from Keplerian rotation, $(v_{\phi}-v_{\rm
       Kep})/|v_{\phi}|$ for model VERT-TRAP. The plot shows the time
     evolution of radial profile computed from a vertical and
     azimuthal averaging.}
    \label{fig:vphi_rad}
  \end{center}
\end{figure}

The marked difference brought about by magnetic field trapping is
highlighted by a comparison of the time evolution of the vertical disk
structure for models VERT and VERT-TRAP in Figs.~\ref{fig:vert_time}
and \ref{fig:vert_time_vt}, respectively. Quasi-periodic oscillations
are seen for model VERT-TRAP, but only at relatively early times in
the run ($t\ltsimm 10\;P^{\rm orb}_{30}$). For the remainder of the
run the quasi-periodic oscillations cease, continuing with sporadic
fluctuations. A possible explanation is that the suppression of the
dynamo cycles by the onset of a strong ($\beta \ltsimm 1$) magnetic
field, following which a mean toroidal magnetic field develops - for
further discussion of this point see \cite{Bai:2013}. Strong turbulent
gas velocities arise at all heights with close-to-transonic streams of
gas emanating from the disk mid plane and rising into the corona.

As the turbulent activity in the disk rises, patches of the disk are
seen to rotate at a sub-Keplerian rate. In Fig.~\ref{fig:vphi_rad} we
show the radial profile of the fractional deviation from Keplerian
rotation as a function of time, where values have been vertically and
azimuthally averaged. One sees that the initial rotation profile
($t=0$) is close to Keplerian. As turbulence establishes in the disk
its rotation deviates from Keplerian, typically by $2\%$ during the
first $10\;P^{\rm orb}_{30}$. At later times, however, considerably
more sub-Keplerian rotation is observed, and during the latter half of
the simulation deviations of $4-10\%$ are seen in the radial range
$10<r<20$. Interestingly, this coincides with the build-up of a strong
magnetic field in the disk (see, e.g., Figs.~\ref{fig:beta} and
\ref{fig:vert_time_vt}), indicating that the additional magnetic
pressure support provided by the strong field alters the disk rotation
rate. Consequently, the shear rate in the disk is modified, which has
implications for the growth rate of the MRI \citep{BH98}. Also
apparent from Fig.~\ref{fig:vphi_rad} is the development of
sub-Keplerian rotation from smaller to larger radii, consistent with
the magnetic field piling-up at the inner radial disk edge in model
VERT-TRAP.

Further differences between models VERT and VERT-TRAP are apparent
from an inspection of the time-averaged vertical profiles of various
quantities shown in Fig.~\ref{fig:vert}. The density profile for model
VERT-TRAP shows a similar rounded shape close to the mid plane but the
profile is much flatter between $1<|z/H|<2$. Both the magnetic field
strength and turbulent velocities are noticeably larger for
VERT-TRAP. The magnetic field is dominated by the toroidal component,
and the strength of the magnetic field drops off with increasing
height away from the mid plane. Similarly, the turbulent velocities
fall-off with height for VERT-TRAP, however at a slower rate that in
the other models.

Although the radial velocity at the inner radial disk edge is reduced
in VERT-TRAP compared to VERT, the radial mass flux for VERT-TRAP is
notably larger, with $\dot{M}_{\rm rad} = 260$ and $159 \ts
P_{30}^{\rm orb-1}$, respectively - see Table~\ref{tab:models}. This
indicates that the density of the accretion flow has increased for
VERT-TRAP. At first sight this result appears at odds with the lower
normalisation of the vertical density profile for VERT-TRAP compared
to VERT in Fig.~\ref{fig:vert}. On inspection of the simulation data,
one sees that mass builds up at the inner radial edge of the disk,
which is outside of the region averaged-over to construct the vertical
profile. Interestingly, model VERT-TRAP displays a considerably larger
vertical mass flux than model VERT ($\dot{M}_{\rm vert}$). In fact,
vertical and radial mass fluxes differ by less than a factor of two
for model VERT-TRAP.  Interestingly, this suggests that the launching
of a wind in a dynamically evolving turbulent disk could be instigated
by a slowing of the radial accretion flow.

\begin{figure}
  \begin{center}
    \begin{tabular}{c}
\resizebox{80mm}{!}{\includegraphics{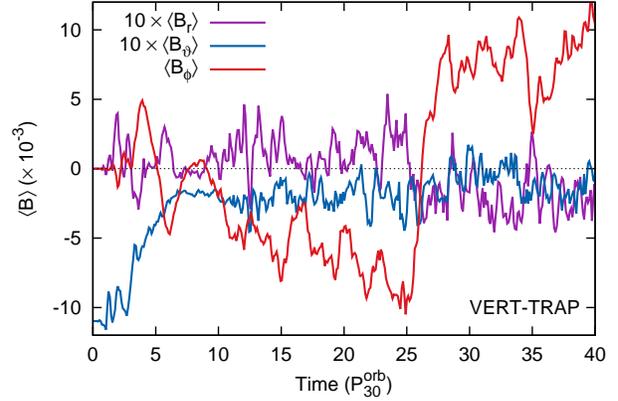}} \\
   \end{tabular}
   \caption{Same as Fig.~\ref{fig:Bav} except for model VERT-TRAP.}
    \label{fig:Bav_vert_vbc}
  \end{center}
\end{figure}

The evolution of mean magnetic fields proceeds differently for model
VERT-TRAP compared to VERT (see Figs.~\ref{fig:Bav} and
\ref{fig:Bav_vert_vbc}). As previously mentioned, the mean toroidal
field grows considerably up until $t\simeq 25 \ts P^{\rm orb}_{30}$
for VERT-TRAP, at which point a sudden sharp field reversal occurs
with a corresponding spike in $\lb \alpha_{\rm P} \rb$. Inspecting the
correlation between vertical field ($| \lb B_{\theta} \rb|$) and
turbulent stresses ($\lb \alpha_{\rm P} \rb$) in
Fig.~\ref{fig:alphaP_Btheta}, one sees that during the initial
transient phase of the simulation ($2<t<10 \ts P^{\rm orb}_{30}$)
similar behaviour to model VERT is observed. However, at late times
($t>20 \ts P^{\rm orb}_{30}$), a much stronger correlation between $|
\lb B_{\theta} \rb|$ and $\lb \alpha_{\rm P} \rb$ is apparent.

In summary, trapping of field in a turbulent disk leads to a highly
magnetized state exhibiting large, highly variable accretion
stresses. In contrast, model TOR-$\beta1$ shows that an initially
equipartition strength field will disperse if outflow is
unconstrained. In this regard the results of model VERT-TRAP and
TOR-$\beta1$ are complementary to the study of efficiently accreting
($\lb \alpha_{\rm P} \rb \sim 0.1$), strongly magnetized ($\beta\sim
1$) disks by \cite{Johansen:2008} and \cite{Gaburov:2012}. Models
VERT-TRAP and TOR-$\beta1$ suggest that a strongly magnetized state
can only arise, and be maintained, when radial flow is
constrained. This point is discussed further in
\S~\ref{subsec:environment}.

\begin{figure}
  \begin{center}
    \begin{tabular}{c}
\resizebox{80mm}{!}{\includegraphics{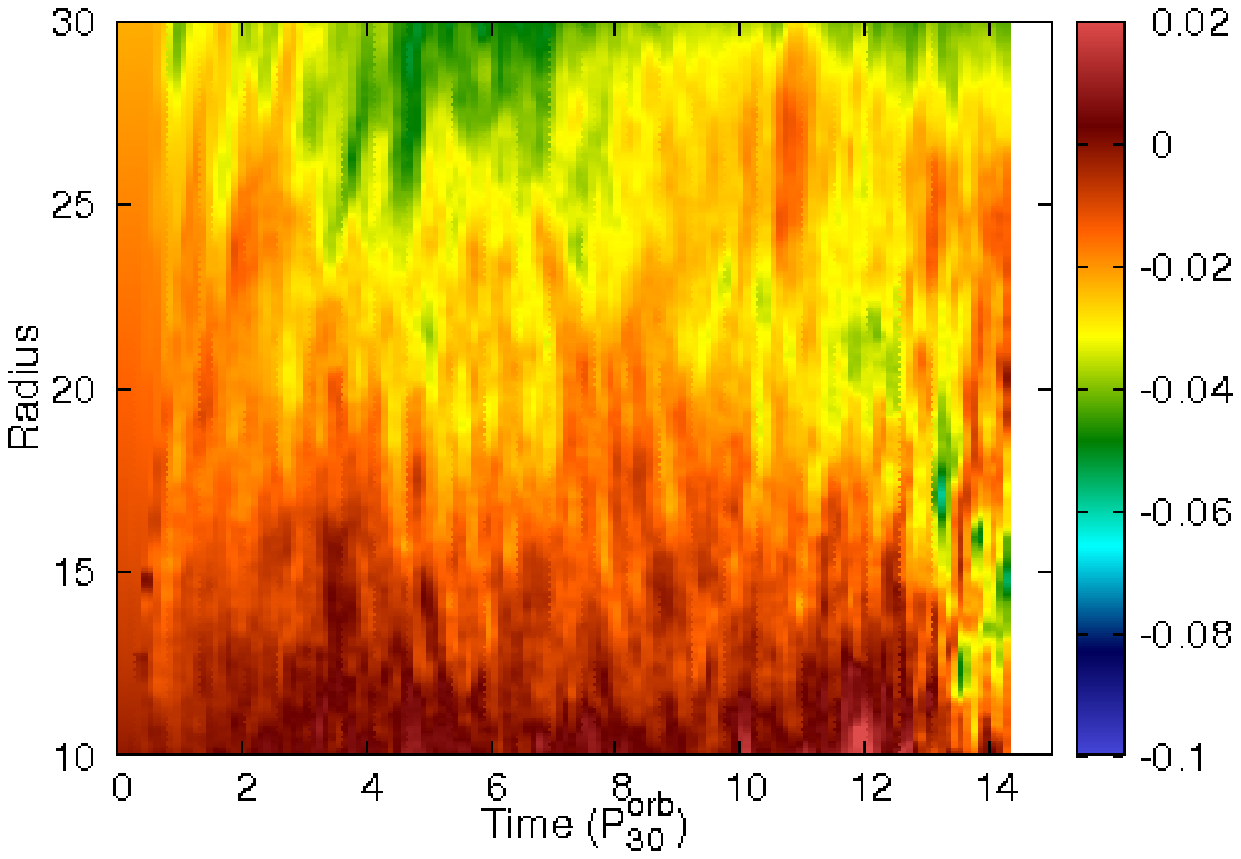}} \\
   \end{tabular}
   \caption{Same as Fig.~\ref{fig:vphi_rad} except for model VERT-B+.}
    \label{fig:vphi_rad_vert-B+}
  \end{center}
\end{figure}

\begin{figure}
  \begin{center}
    \begin{tabular}{c}
\resizebox{80mm}{!}{\includegraphics{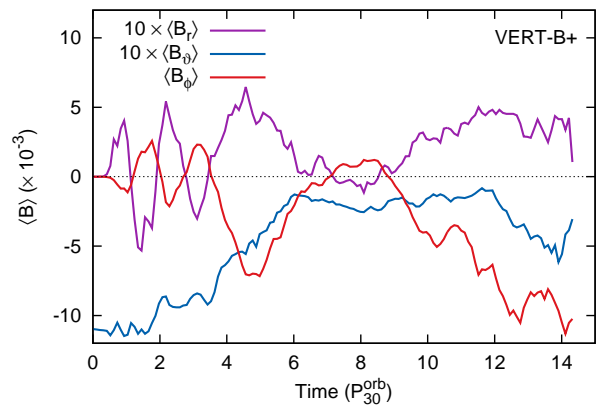}} \\
   \end{tabular}
   \caption{Same as Fig.~\ref{fig:Bav} except for model VERT-B+.}
    \label{fig:Bav_vert_B+}
  \end{center}
\end{figure}

\subsection{Injecting vertical magnetic field (model VERT-B+)}
\label{subsec:vert-Bthadd}

A general trend from the models in \S~\ref{subsec:field_topology} is
that during the initial transient phase of disk evolution the field
configuration is disrupted and not entirely conserved. For example,
the initially purely vertical magnetic field in model VERT leaves the
domain in the accretion flow and its subsequent evolution becomes
similar to other models initialised with different field topologies
(e.g. TOR-$\beta20$ and LOOP). In this section we use model VERT-B+ to
examine the disk evolution that occurs when a weak vertical magnetic
field is steadily added to the simulation domain, replenishing the
field that is advected away at early times. The motivation for model
VERT-B+ comes from the possibility that in astrophysical disks
additional magnetic field (i.e. beyond that contained in the
simulation initial conditions in this case) may flow into the disk
from larger radii or, alternatively, a field component on scales much
larger than the simulation domain may pervade the disk.

The setup for model VERT-B+ is identical to that of model VERT, with
the exception that a constant $B_{\theta}$ is added to the simulation
domain between $12<r<30$ and for all $\theta$ and $\phi$. The
$\theta$-component of the magnetic field is added at a constant rate
such that it would generate a $\beta=2000$ strength field in a time
interval of $40\;P^{\rm orb}_{30}$. Note that the injected
$B_{\theta}$ is divergence-free and is added to the staggered mesh
component of $B_{\theta}$.

The effect of injecting a weak vertical magnetic field on the
turbulent stresses is shown in Fig.~\ref{fig:alpha_tests}. Up until a
simulation time of $\simeq 10\;P^{\rm orb}_{30}$ the $\lb \alpha_{\rm
  P} \rb$ values for models VERT-B+ and VERT are quite
similar. However, from $t\sim 11\;P^{\rm orb}_{30}$ onwards the models
diverge and the turbulent stress in model VERT-B+ rises sharply. By
$t= 14.4\;P^{\rm orb}_{30}$ the stress reaches $\lb \alpha_{\rm P} \rb
\simeq 0.25$ - we were unable to evolve model VERT-B+ beyond this
point as the highly turbulent state of the disk proved problematic for
the simulation code. As such, VERT-B+ does not reach a quasi-steady
state making an analysis of time-averaged quantities/profiles
difficult. From the short duration of VERT-B+ it is unclear whether additional
vertical mass-loss in the form of a wind is triggered by the burst in
activity at $t\gtsimm 11\;P^{\rm orb}_{30}$.

During the simulation the disk rotation deviates from Keplerian
rotation (Fig.~\ref{fig:vphi_rad_vert-B+}). Patches of sub-Keplerian
rotation are observed between $20<r<30$ at times later than $5\;P^{\rm
  orb}_{30}$. However, at $t\gtsimm 10 \;P^{\rm orb}_{30}$ regions of
sub-Keplerian rotation also begin to extend to smaller radii,
coincident with the onset of the accretion burst. Recalling the
results of model VERT-TRAP where sub-Keplerian rotation was coincident
with strong magnetic fields in the disk, a possible explanation for
the emergence of patches of sub-Keplerian rotation in model VERT-B+ is
that strong vertical field is dragged through the disk by the
accretion flow. This causes a back-reaction, with further field growth
leading to an increase in magnetic field, as in the classic picture of
field amplification via the MRI \citep{BH92}.

The contrast in model evolution between VERT-B+ and VERT in the time
interval $11 \ltsimm t \ltsimm 14\;P^{\rm orb}_{30}$ does, however,
highlight important differences resulting from vertical field
replenishment. Firstly, it is worthwhile to note that the rate of
injection of $B_{\theta}$ into the domain is not abrupt, yet the
accumulation of additional field is sufficient to bring about a burst
in accretion. During this burst phase the strength of the radial,
vertical, and toroidal magnetic field components are all seen to grow,
illustrated by a decrease the respective plasma-$\beta$ values
(Fig.~\ref{fig:beta}). In essence, only a relatively weak vertical
magnetic field need be injected into the disk to cause a considerable
change in the accretion activity of the disk. Local shearing-box
models by \cite{Bai:2013} display similar behaviour whereby the
injection of a vertical magnetic field into an already turbulent disk
causes considerable changes to the turbulent state. Model VERT-B+
extends this picture to global disks, albeit with a simple
prescription for magnetic field injection. A detailed study of the
parameter space relating to field injection (e.g., rate of field
injection, topology of injection field, radius of field injection)
would require additional models that are beyond the scope of the
current work.

The cause of the transient outburst clearly must be related to the
injection of vertical magnetic field. However, on inspection, the
exact details are subtle. Firstly, if allowed to accumulate, the
injected magnetic field would equate to a $\beta=2000$ vertical field
in a time span of $40\;P^{\rm orb}_{30}$. At $z=0$ one finds from
Eq~(\ref{eqn:lambdaMRI}) that $\Lambda_{\rm MRI-z}/\Delta z \approx
8.89 n_{\rm z-H} \beta_{z}^{-1/2}$, where $\Delta z$ and $n_{\rm z-H}$
are the cell size and number of cells per scale-height in the
$z$-direction, respectively. For $\beta_{z}=2000$ one finds
$\Lambda_{\rm MRI-z}/\Delta z = 8.8$, indicating adequate resolution
of the fastest growing MRI mode for the initial field in models VERT,
VERT-TRAP, and VERT-B+. However, the transient arises after $\simeq
11\;P^{\rm orb}_{30}$ in model VERT-B+, after which time the
accumulated field would have $\Lambda_{\rm MRI-z}/\Delta z \sim 2.4$,
which is borderline in terms of adequate resolution for the MRI to
develop. The next clue comes from examining the mean field evolution
and plasma-$\beta$ for model VERT-B+. The mean vertical field in model
VERT-B+ levels-off at $t\simeq 6\;P^{\rm orb}_{30}$ (see
Fig.~\ref{fig:Bav_vert_B+}), in contrast to the continuing decline
seen for VERT (Fig.~\ref{fig:Bav}). Then, at $t \simeq 11 \;P^{\rm
  orb}_{30}$, $| \lb B_{\theta} \rb |$ grows, which directly coincides
with the burst in accretion activity (Fig.~\ref{fig:alpha_tests}). The
magnitude of the vertical field, measured by the vertical field
plasma-$\beta$, also grows at this time (illustrated by a decrease in
$\beta_{\theta}$ in Fig.~\ref{fig:beta}). Therefore, although the
accumulated field is relatively weak in terms of adequate resolution
it counters the loss of field in the accretion flow to the extent that
the disk evolution is altered. This may arise from the injected field
building-up in dormant areas of the disk (e.g. at larger radii where
the MRI takes longer to set-in initially) which, upon becoming
actively turbulent, triggers a burst in accretion.

\subsection{Transient accretion burst via mass injection (model TOR-M+)}
\label{subsec:tor-inj}

The simulated disks in models TOR-$\beta1$, TOR-$\beta20$, VERT, and
LOOP evolve to a similar quasi-steady turbulent state with no
indication of a large amplitude recurrent burst of
accretion\footnote{\cite{Flock:2012} argue that simulations using an
  azimuthal extent of $< \pi$ will be subject to recurring accretion
  bursts, related to an artificial build-up of low azimuthal
  wavenumber energy triggering a secondary phase of linear
  non-axisymmetric MRI growth. However, the simulations presented in
  this work use an azimuthal extent of $\pi/2$ with no sign of a
  recurrent linear-MRI phase, suggesting that run times longer than we
  have considered may be required for sufficient energy to build
  up.}. Astrophysical disks, on the other hand, do exhibit recurring
transient bursts. This variability is often attributed to thermal
instability \citep{Meyer:1981, Pringle:1981, Faulkner:1983}, whereby
the build-up of matter during the quiescent state leads to a sudden
increase in disk opacity related to the ionisation of
hydrogen. Subsequently, the disk heats up and the turbulent activity
rises. The compatibility of this scenario with magnetorotational
turbulence has recently been demonstrated by \cite{Latter:2012} and
\cite{Hirose:2014}. From a slightly different tack, \cite{Gammie:1998}
suggest that dwarf novae eruptions originate from a combination of
magnetorotational turbulence with some secondary episodic (global)
instability \citep[e.g.][]{Papaloizou:1984}.

Common to the above suggestions for triggering a transient burst is
the build-up of mass in the disk. In this section we examine the
results of a supplementary simulation (model TOR-M+) which tests the
response of a turbulent magnetic global disk to the rapid addition of
zero angular momentum mass. This model uses the state of TOR-$\beta20$
at $t=25 \ts P^{\rm orb}_{30}$ for its initial conditions. Then, in
the interval $25<t<26 \ts P^{\rm orb}_{30}$ the mass source term
(described in \S~\ref{subsec:hydromodel}) is modified to: i) relax the
density towards twice that of the initial conditions, and, ii)
conserve angular momentum in density increments, $\delta \rho $, such
that the azimuthal velocity, $v_{\phi}$, is respectively modified by
$\delta v_{\phi} = -v_{\phi}\delta \rho / (\rho + \delta \rho)$. We
note that model TOR-M+ is also applicable for understanding the
subsequent disk evolution following stochastic accretion in the disk
tearing scenario proposed by \cite{Nixon:2012, Nixon:2013}.

Shortly after the enhanced mass injection commences the disk undergoes
a transient burst of accretion activity and a sharp rise in $\lb
\alpha_{\rm P} \rb$ for model TOR-M+ at $t\simeq 25 \ts P^{\rm
  orb}_{30}$ (Fig.~\ref{fig:alpha_tests}). $\lb \alpha_{\rm P} \rb$
subsequently decays, with a mild recurrence at $t\simeq 40 \ts P^{\rm
  orb}_{30}$, possibly related to clumps of mass compressing magnetic
field as they advect through the disk. This premise is supported by
$\beta_{\rm tot}$ falling to $\simeq 7$ at this time (compared to
$\beta_{\rm tot} \simeq 18$ for model TOR-$\beta20$) - see
Fig.~\ref{fig:beta} and Table~\ref{tab:models}. Examining the vertical
structure of the disk in model TOR-M+ (see Fig.~\ref{fig:vert}) it
bears close similarity to model TOR. The most notable difference is in
the strength of the magnetic field, which is clearly stronger in all
components for model TOR-M+ compared to TOR, but more apparently so
for the radial and vertical field.

The mechanism that triggers the outburst is two-fold. Firstly,
increasing the density reduces the Alfv{\' e}n speed, which in turn
reduces the wavelength of the fastest growing MRI mode (see
Eq~\ref{eqn:lambdaMRI}). This opens up additional fast growing, short
wavelength MRI modes which can inject energy on the smallest
scales. Secondly, reducing the azimuthal velocity in regions where
mass is added causes gravitational attraction to exceed the
centrifugal force. Hence, over-dense regions move radially inwards,
dragging field lines with them which are then sheared. The result is a
rise in radial field followed by conversion to toroidal field, both of
which act to enhance the turbulent stresses.

\section{Discussion}
\label{sec:discussion}

\subsection{How big is $\alpha$?}

The stresses produced by magnetorotational turbulence effectively
transport angular momentum. However questions have been raised about
whether the level of angular momentum transport produced by MRI-driven
turbulence (quantified by $\lb \alpha_{\rm P} \rb$) is actually
sufficient to explain observations of astrophysical accretion
disks. For example, \cite{King:2007} compiled results from
shearing-box studies and highlighted a disparency between simulated
$\lb \alpha_{\rm P} \rb$ values of $ \ltsimm 0.01$ and observationally
inferred values of $\lb \alpha_{\rm P} \rb \simeq 0.1-0.3$ for
post-outburst dwarf novae disks.

There are some important points to note about this
discrepancy. Firstly, unstratified shearing-box studies with a net
vertical field display $\lb \alpha_{\rm P} \rb$ values as high as
0.2-0.7 \citep[][]{Hawley:1995, Sano:2004, Simon:2009}. However, a
typical consequence of a net vertical field is a disk wind, whereas a
number of the objects discussed by \cite{King:2007} show no evidence
of a wind, jet, or large-scale vertical field of any description.

The evolution of a global disk threaded with a net vertical field has
been explored in this work - see also \cite{Beckwith:2009} and
\cite{Suzuki:2014}. Interestingly, during the transient phase of model
TOR-$\beta1$, $\lb \alpha_{\rm P} \rb$ reaches values as high as $\sim
0.37$, with no associated disk wind (i.e. the vertical-to-radial mass
flux ratio is small). At late-times in the simulation the net vertical
field has been effectively removed, slipping away in the accretion
flow, and irrespective of the initial field configuration one finds
$\lb \alpha_{\rm P} \rb \simeq 0.032-0.036$. A similar range of values
for $\lb \alpha_{\rm P} \rb $ is seen in previous global disk
simulations. For example, \cite{Suzuki:2014} find $\lb \alpha_{\rm P}
\rb\sim 0.01-0.02$ during the late-time evolution, whereas the study
by \cite{Hawley:2013} (for which the models are not evolved beyond the
initial transient phase) reports $\lb \alpha_{\rm P} \rb $ as high as
0.15. Taking stock of the above results, there is an obvious
suggestion that magnetorotational turbulence can produce sufficiently
large $\lb \alpha_{\rm P} \rb $ values to gain agreement with
observation. However, the disk must either be made to undergo a
transient burst, or, a net vertical field must be retained/replenished
over dynamically significant timescales. We have demonstrated the
viability of these channels with models TOR-M+, VERT-TRAP, and
VERT-B+, which produce $\lb \alpha_{\rm P} \rb \simeq 0.1- 0.25$.

\subsection{External/environmental influence}
\label{subsec:environment}

An isolated disk will evolve to a common quasi-steady turbulent state,
irrespective of the initial conditions and/or initial disturbance to
the disk. This is illustrated by models TOR-$\beta20$, TOR-$\beta1$,
VERT, and LOOP, for which statistical measures of the turbulence are
similar during the late-time disk evolution \citep[see also][where the
influence of different initial perturbations is
examined]{Parkin:2013}. As discussed in the previous section,
astrophysical disks do exhibit transient evolution and in some cases
changes in the accretion state. Model VERT-TRAP shows that altering
the rate that mass and magnetic field can leave the disk can cause
abrupt changes to the accretion state. In essence, events occurring
exterior to the disk can directly influence the accretion state by
regulating the inflow/outflow of material/field.

All of this points to the importance of the environment outside of the
disk. One must then give consideration to the timescale(s) associated
with the external process(es). For example, if the disk expels
magnetic flux in the accretion flow at a rate faster than it can be
digested by the central object, be it a star or a black hole, then the
build-up of magnetic flux may impinge on the disk
\citep[e.g.][]{Igumenshchev:2003, Romanova:2012}. Furthermore,
accretion of magnetized gas onto the disk at a rate faster than it can
be processed by the turbulence can lead to a strongly magnetized state
\citep{Gaburov:2012}.

The environment surrounding the accretion disk may enforce
preservation of field topology, or provide a source of poloidal
magnetic flux into the disk over a dynamically important
timescale. For example, the conservation of magnetic flux in collapse
scenarios \citep{Bisnovatyi-Kogan:1974, Bisnovatyi-Kogan:1976,
  Mouschovias:1976}, accretion of magnetic field in a galactic nucleus
\citep{Proga:2003}, or wind accretion in a binary system
\citep{Frank:2002}. Global models similar to those presented in this
study could be readily augmented to examine these scenarios.

\subsection{Wind launching}

Launching a wind from a turbulent magnetized disk remains a
challenging problem. Recent studies adopting shearing-box models have
demonstrated that vertical outflows arise from a patch of a turbulent
disk if initialised with a net flux vertical field
\citep{Suzuki:2009,Ogilvie:2012,Moll:2012, Lesur:2013, Fromang:2013,
  Bai:2013}. However, the radially periodic shearing-box boundary
conditions preserve the net vertical field topology (see
\S~\ref{sec:mean_field}). In contrast, the vertical field in model
VERT exits the disk in the accretion flow \citep[see
also][]{Suzuki:2014} and during the quasi-steady state the ratio of
vertical-to-radial mass flux is $\sim 11\%$. Model VERT-TRAP, on the
other hand, shows that if the radial flow out of the disk is
constrained (so as to trap vertical and azimuthal field) then the
vertical-to-radial mass flux reaches $\sim59\%$. Hence, the results
indicate that time-steady disks do not launch winds if the radial flow
is unconstrained, whereas wind-launching may be triggered if the rate
at which vertical/azimuthal field exits the disk is impeded (compared
to the steady-state rate).

\section{Conclusions}
\label{sec:conclusions}

The dependence of turbulent stresses on initial field configuration,
magnetic field trapping due to inhibited radial outflow, magnetic
field injection, and rapid mass injection has been investigated using
global MHD disk simulations. Properties of the saturation stress level
during the early-time transient phase of disk evolution show a
dependence on the initial magnetic field. This finding is consistent
with models presented by \cite{Hawley:2013} which focused on the
transient phase during roughly the first $\simeq 10.3 \ts P^{\rm
  orb}_{30}$ of the simulation. However, a major new result in this
work, and one which was achieved by utilising a mass source term to
avoid the exhaustion of disk mass by ongoing accretion, is that
irrespective of the strength and topology of the initial magnetic
field an {\it isolated} magnetorotationally turbulent disk evolves to
a statistically similar quasi-steady state that is characterised by:
i) a volume-averaged stress-to-gas-pressure ratio, $\lb \alpha_{\rm P}
\rb \simeq 0.032-0.036$, ii) a gas-to-magnetic pressure ratio, $\beta
\simeq 16-18$, and, iii) zero-net time-averaged radial and azimuthal
magnetic fields.

Simulations exploring the influence of trapping of magnetic field,
vertical field injection, and of rapid mass injection, on the
turbulence reveal that these mechanisms lead to relatively large
stresses in the disk. Trapping of magnetic field produces a strongly
magnetized disk with $\lb \alpha_{\rm P} \rb \sim 0.2$ and
considerable vertical mass flux in the form of an intermittent
wind. The injection of a vertical magnetic field, even a relatively
weak one that equates to an accumulation of a $\beta=2000$ field over
a time interval of $40\;P^{\rm orb}_{30}$, is sufficient to trigger a
burst in accretion with stresses reaching $\lb \alpha_{\rm P} \rb \sim
0.25$. Rapid mass injection leads to a sudden sharp increase in
turbulent stresses, peaking at $\lb \alpha_{\rm P} \rb \sim 0.1$ and
subsequently decaying to the pre-outburst level. The results of these
simulations may be relevant for understanding accretion state changes,
transient bursts, and large observationally inferred accretion
efficiencies for astrophysical disks. Indeed limit cycle behaviour
reminiscent of dwarf novae eruptions could be produced with the
inclusion of a temperature (and perhaps also surface density)
dependent resistivity, following similar suggestions by
\cite{Balbus:2008b}, \cite{Lesaffre:2009}, and \cite{Latter:2012}.

\subsection*{Acknowledgements}
I thank Geoffrey Bicknell, Yuri Levin, and Daniel Price for helpful
discussions, the Australian Research Council's Discovery Projects
scheme (project number DP1096417) for financial support, and the
National Computational Infrastructure for computational resources
through access to the Raijin supercomputer.


 \newcommand{\noop}[1]{}

\label{lastpage}

\end{document}